\documentclass[aps,prstab,preprint,groupedaddress, amsmath,showpacs, preprintnumbers]{revtex4}
\usepackage{graphicx}% Include figure files
\usepackage{dcolumn}% Align table columns on decimal point
\usepackage{bm}% bold math
\usepackage{psfrag}

\begin{document}

% Use the \preprint command to place your local institutional report
% number in the upper righthand corner of the title page in preprint mode.
% Multiple \preprint commands are allowed.
% Use the 'preprintnumbers' class option to override journal defaults
% to display numbers if necessary
%\preprint{}
\preprint{\vtop{\hbox{SLAC-PUB-12136}}}

%Title of paper
\title{Matrix formalism of synchrobetatron coupling}

% repeat the \author .. \affiliation  etc. as needed
% \email, \thanks, \homepage, \altaffiliation all apply to the current
% author. Explanatory text should go in the []'s, actual e-mail
% address or url should go in the {}'s for \email and \homepage.
% Please use the appropriate macro foreach each type of information

% \affiliation command applies to all authors since the last
% \affiliation command. The \affiliation command should follow the
% other information
% \affiliation can be followed by \email, \homepage, \thanks as well.
\author{Xiaobiao Huang}
\email[]{xiahuang@slac.stanford.edu}
%\homepage[]{Your web page}
\thanks{}
%\altaffiliation{}
\affiliation{Stanford Linear Accelerator Center, Menlo Park, CA 94025}

%Collaboration name if desired (requires use of superscriptaddress
%option in \documentclass). \noaffiliation is required (may also be
%used with the \author command).
%\collaboration can be followed by \email, \homepage, \thanks as well.
%\collaboration{}
%\noaffiliation

\date{\today}
%\date{January 3, 2007}

\begin{abstract}
In this paper we present a complete linear synchrobetatron coupling formalism by studying the   
transfer matrix which describes linear horizontal and longitudinal motions.  
With the technique established in  the linear horizontal-vertical coupling study
[D. Sagan and D. Rubin, Phys. Rev. ST Accel. Beams {\bf 2}, 074001 (1999)], we 
found a transformation to block diagonalize the transfer matrix and decouple the 
betatron motion and the synchrotron motion. By separating the usual dispersion term 
from the horizontal coordinate first, we were able to obtain analytic expressions 
of the transformation under reasonable approximations.  
We also obtained the perturbations to the betatron tune and the Courant-Snyder functions. 
The closed orbit changes due to finite energy gains at rf cavities and 
radiation energy losses were studied by  
the 5$\times$5 extended transfer matrix with the fifth column describing kicks in the 4-dimension 
phase space. 

\end{abstract}

% insert suggested PACS numbers in braces on next line
% insert suggested keywords - APS authors don't need to do this
%\keywords{}

%\maketitle must follow title, authors, abstract, \pacs, and \keywords
\maketitle

% body of paper here - Use proper section commands
% References should be done using the \cite, \ref, and \label commands
\section{Introduction \label{secIntro} }
% Put \label in argument of \section for cross-referencing
%\section{\label{}}

The synchrobetatron coupling (SBC) comes from dispersion at rf cavities 
and the path length dependence on the amplitude of betatron motion. 
The dispersion at an rf cavity makes the longitudinal kicks received from 
the cavity affect the 
betatron motion. Since the longitudinal kicks depend on the arrival time of the particles, 
 the longitudinal motion is coupled to the betatron 
motion. On the other hand, particles with different betatron amplitudes have 
different path lengths which affect the arrival time. So betatron motion is 
also coupled to the longitudinal motion. 

Traditionally many authors treated SBC with the Hamiltonian dynamics approach~\cite{SuzukiPA,Corsten}, which 
is a general and complete description and naturally covers effects of nonlinearities. 
It is very useful for the study of synchrobetatron resonances since in such cases one can focus on 
only the resonant term of the synchrobetatron 
potential. However, the Hamiltonian approach is cumbersome for the off-resonance 
cases which are most common for storage ring operations. 
In the linear case, a parallel approach is the matrix formalism first proposed by Chao in 
Ref.~\cite{AChao79} which described the construction of the $6\times6$ transfer matrices and 
the decomposition of the coupled motion to the eigen-modes of the one-turn transfer matrix. 
Ref.~\cite{AChao79} also described an iterative procedure to include the nonlinear effects. 
Chao later applied the method to compute the beam tilt angle between $x$-$z$ plane due to 
rf cavities~\cite{AChaoXZTilt}. 

Recently the study of low-alpha lattices stimulated Shoji's work on the path length effect 
which yielded an important result of bunch lengthening due to betatron emittance and dispersion~\cite{ShojiBunLen}. 
On the other hand Ref.~\cite{GuoXL} studied the SBC-induced closed orbit change by   
considering the dipole-like kicks in the horizontal betatron phase space due to 
the sudden changes of energy at a nonzero-dispersion rf cavity. 
The authors derived the horizontal closed orbit changes induced by the finite energy gains at the rf cavities 
and verified with both simulations and experiments. 

In this paper, we will study the linear synchrobetatron coupling 
under the transfer matrix framework without considering diffusion and damping 
due to radiation. 
Since the vertical motion is not essential to the SBC, we don't consider it 
for simplicity reasons.
We then study the 4$\times$4 horizontal-longitudinal transfer matrix 
in the same manner as the horizontal-vertical coupling is studied~\cite{SaganPRSTAB}. 
Namely, 
we try to decouple the horizontal and longitudinal motions by using a coordinate transformation 
to block diagonalize the transfer matrix and obtain the normal modes, in this 
case, the pure betatron mode and the pure synchrotron mode. 
We first study the fixed-energy case in which no rf cavity exists (or the rf gap voltage 
is set to zero). The transformed coordinates include the usual betatron coordinates and 
momentum deviation coordinate. 
But the longitudinal phase coordinate is modified by a term involving $D x'_\beta-D' x_\beta$ 
which corresponds to the bunch lengthening effect studied by Shoji~\cite{ShojiBunLen}.  
In cases with rf cavities, we first apply the previous fixed energy transformation to 
separate the dispersion term. Since the synchrotron motion 
is usually slow, the coupling (off-diagonal) blocks  of the transfer matrix for the new 
coordinates are small. Therefore we 
can perform the block diagonalization procedure  
proposed in Ref.~\cite{SaganPRSTAB} approximately yet with high precision. 
The transformation matrix is expressed analytically with the usual parameters such as 
the Courant-Snyder parameters, dispersion 
functions and the rf voltage slope. When the normal modes are obtained, we can  calculate 
their contributions to the beam width and bunch length. 
Perturbations to horizontal betatron motion due to SBC, including changes to the betatron 
tune and Courant-Snyder functions are also obtained. 

It is well known that SBC causes changes to the beam orbit~\cite{JMaidment, VossNote, AChaoOrbit2, MJLee75, GuoXL}. 
In this study we find the closed 
orbit in the 4-dimension phase space in an analytical form through 
the 5$\times$5 extended transfer matrix method~\cite{AChao79}, with the fifth column containing the 4-dimension 
kicks the beam receives from rf cavities and dipole magnets. 
Both finite energy gains at rf cavities and radiation energy losses in bending dipoles are considered.  
The radiation energy loss is random by nature. However, since usually hundreds of photons are emitted 
in one revolution, much more than the number of dipole magnets in which emission happens, 
we consider the radiation energy loss as a steady and uniform process. 
The radiation energy losses contribute additional terms to the horizontal closed orbit change. 
The above results are verified with the accelerator modeling code AT~\cite{AccelTool}. 

This paper is organized as follows. 
Section~\ref{secIntro} is this introduction. 
Section~\ref{secFixedE} describes the block diagonalization of the 4$\times$4 transfer matrix in 
the fixed energy case. 
Section~\ref{secMatForm} presents the matrix formalism of the synchrobetatron coupling. 
Section~\ref{secCOD} is the calculation of closed orbit changes induced by finite energy gains  
at rf cavities and radiation energy losses.  
Section~\ref{secSim} shows simulation results and the comparison to the theory. 
Section~\ref{secDiscu} gives the conclusions.

\section{Block diagonalization for a fixed energy ring \label{secFixedE}}

The 4-dimension coordinate vector is ${\bf X} = ({\bf x}^T, {\bf l}^T)^T$, where 
the horizontal coordinate vector is ${\bf x} = (x, x')^T$ and the longitudinal 
coordinate vector is ${\bf l} = (c\tau, \delta)^T$. The $c\tau$ coordinate instead of the phase 
coordinate  
$\phi= \phi_s - \frac{h}{R}c\tau$, where $\phi_s$ is the synchronous phase, $h$ 
is the harmonic number and $R$ is the average ring radius, 
is used to avoid the appearance of scaling factors $h/R$ in the 
transfer matrix. Note a negative $c\tau$ indicates the particle is behind the 
synchronous particle. The coordinates at the entrance and the exit of an accelerator component are related 
through its transfer matrix $\bf T$ which 
can be divided into 2$\times$2 blocks $\bf M$, $\bf E$, $\bf F$ and $\bf L$ such that 
\begin{eqnarray}\label{eqGeneralM44}
{\bf T} = \left( \begin{array}{cc} {\bf M} & {\bf E} \\ 
{\bf F}   &  {\bf L} \end{array}\right). 
\end{eqnarray}
In general, the coupling  
blocks ${\bf E}$, $\bf F$ of a single component are nonzero only for dipole magnets. 
And for time-independent components   
(which include most common accelerator components except rf components such as rf cavity, rf dipoles and rf quadrupoles), 
the coupling blocks have two zero matrix elements such that 
\begin{eqnarray}\label{eqNonrfEFblock}
{\bf E} = ({\bf 0}, {\bf e}) \qquad {\rm and} \quad
{\bf F} = \left( \begin{array}{c} {\bf f}^T \\  
{\bf 0} \end{array}\right), 
\end{eqnarray}
where ${\bf e}$ and ${\bf f}$ are 2-component column vectors and $\bf 0$'s are  
zero vectors of suitable sizes. The zero elements in matrix ${\bf E}$ and ${\bf F}$ 
are consequences of 
the fact that the horizontal coordinates don't depend on the arrival time of the particles  
and the horizontal coordinates don't cause energy changes in such components.  
The $\bf L$ blocks for rf cavities and other components are 
\begin{eqnarray}
{\bf L}_{\rm rf} &=& \left( \begin{array}{cc} 1 & 0 \\  
w & 1 \end{array}\right) \qquad {\rm and} \qquad 
{\bf L}_{\rm other} = \left( \begin{array}{cc} 1 & \eta \\  
0 & 1 \end{array}\right),  \\
w &=& \frac{e}{E}\frac{dV}{cd\tau}=  -\frac{eV_0\cos\phi_s}{E}\frac{h}{R}, 
\end{eqnarray}
where $V_0$ is the gap voltage and $E$ is the beam energy.  The $\eta$ 
parameter is related to the fractional phase 
slippage factor and is nonzero only for dipole magnets if we assume all particles have the 
same velocity $c$, the speed of light. 

The transfer matrix for an accelerator section is the matrix product of the transfer matrices of the 
sequence of components which it consists of.
For any sequence of components not containing an rf cavity, condition Eq. (\ref{eqNonrfEFblock}) 
still holds. 
The symplecticity requirement of the transfer matrices of such sections 
is equivalent to: $\bf M$ and $\bf L$ are symplectic and 
\begin{eqnarray}\label{eqMJ2fSimpNorf}
{\bf M} {\bf J}_2 {\bf f} = {\bf e}, \qquad 
{\bf J}_2 = \left( \begin{array}{cc} 0 & 1 \\  
-1 & 0 \end{array}\right).  
\end{eqnarray}
The $\bf e$ vector and $\eta$ parameter for a section from point 1 to point 2 can be written in integral 
forms
\begin{eqnarray}
{\bf e}_{21} &=& \int_{s_1}^{s_2} {\bf M}(s_2|s) \left( \begin{array}{c} 
0   \\  \frac{ds}\rho \end{array}\right),  \\
\eta_{21} &=& -\int_{s_1}^{s_2} e^1(s_2|s) \frac{ds}\rho,  
\end{eqnarray}
where $s$ is the arc length along the reference orbit, 
$s_2|s$ means ``from $s$ to $s_2$'' , $\rho$ is the bending radius and 
$e^1$ is the first element of ${\bf e}$. %Note that subscript $21$ indicates ``from point 1 to point 2''.

The one-turn transfer matrix with an rf cavity in the ring can be derived in the following 
way. Suppose the rf cavity is located at point 2 and we want to 
calculate the transfer matrix at an arbitrary point 1. 
The transfer matrices between point 1, 2 and at the rf cavity are 
\begin{eqnarray}
{\bf T}_{21} &=& 
\left( \begin{array}{cc} {\bf M}_{21} & {\bf E}_{21}   \\ 
{\bf F}_{21}   &  {\bf L}_{21} \end{array}\right),      
\quad 
{\bf T}_{12} = 
\left( \begin{array}{cc} {\bf M}_{12} & {\bf E}_{12}  \\ 
{\bf F}_{12}   &  {\bf L}_{12}  \end{array}\right),      
\quad
{\bf T}_{\rm rf} = 
\left( \begin{array}{cc} {\bf I} & {\bf 0}  \\ 
{\bf 0}   &  {\bf L}_{\rm rf}  \end{array}\right),      
\end{eqnarray}
where ${\bf M}_{21}$ and ${\bf M}_{12}$ are the horizontal transfer matrices between 
the points, $\bf I$ is the 2$\times$2 identity matrix and 
${\bf L}_{21}$ and ${\bf L}_{12}$ differ from the 2$\times$2 identity matrix 
by their $(1,2)$ elements of $\eta_{21}$ and $\eta_{12}$, respectively. 
The one turn transfer matrix at point 1 is 
\begin{eqnarray}
{\bf T} &=& {\bf T}_{12}{\bf T}_{\rm rf}{\bf T}_{21} = \left( \begin{array}{cc} {\bf M}_1 & {\bf E}_1 \\ 
{\bf F}_1   &  {\bf L}_1 \end{array}\right),
\end{eqnarray}
which can be expressed as 
\begin{eqnarray}\label{eqTGeneral}
{\bf T} &=&  \left( \begin{array}{cc} {\bf M}_{12}{\bf M}_{21}+{\bf E}_{12}{\bf L}_{\rm rf}{\bf F}_{21} & 
{\bf M}_{12}{\bf E}_{21}+{\bf E}_{12}{\bf L}_{\rm rf}{\bf L}_{21} \\ 
{\bf F}_{12}{\bf M}_{21}+{\bf L}_{12}{\bf L}_{\rm rf}{\bf F}_{21}   &  
{\bf F}_{12}{\bf E}_{21}+{\bf L}_{12}{\bf L}_{\rm rf}{\bf L}_{21} \end{array}\right).
\end{eqnarray}

We first consider the case when there is no rf cavity or the cavity is turned off. This corresponds to 
$w=0$ and ${\bf L}_{\rm rf} = {\bf I}$. Consequently ${\bf E}_1$ and ${\bf F}_1$ satisfy conditions 
Eqs. (\ref{eqNonrfEFblock},\ref{eqMJ2fSimpNorf}). We intend to introduce a transformation 
${\bf X} = {\bf U} {\bf X}_n$
to block diagonalize the new transfer matrix ${\bf T}_n={\bf U}^{-1}{\bf TU}$. 
It can be shown that this is achieved by 
\begin{eqnarray}
{\bf U} &=& 
\left( \begin{array}{cc} {\bf I} & {\bf D}_1  \\ 
-{\bf D}_1^+   &  {\bf I} \end{array}\right), \qquad
{\bf U}^{-1} = 
\left( \begin{array}{cc} {\bf I} & -{\bf D}_1  \\ 
{\bf D}_1^+   &  {\bf I} \end{array}\right),
\end{eqnarray}
where 
\begin{eqnarray}\label{eqdefDisp}
{\bf D}_1 &=& ({\bf I}-{\bf M}_1)^{-1} {\bf E}_1 
\end{eqnarray}
and ${\bf D}_1^+$ is its symplectic conjugate~\cite{SaganPRSTAB}.  
We may introduce a column vector ${\bf d}_1=(D_1, D'_1)^T$ 
so that ${\bf D}_1=({\bf 0}, {\bf d}_1)$.
Eq. (\ref{eqdefDisp}) defines the dispersion functions $D_1$ and $D'_1$ at point 1. 
One can also show that 
\begin{eqnarray}
{\bf E}_{21} &=& {\bf D}_2 - {\bf M}_{21} {\bf D}_1
\end{eqnarray}
from ${\bf T}_2{\bf T}_{21} = {\bf T}_{21}{\bf T}_1$.   
The new transfer matrix is found to be 
\begin{eqnarray}
{\bf T}_n &=& 
\left( \begin{array}{cc} {\bf M}_1 & {\bf 0}  \\ 
{\bf 0}  &  {\bf L}_{1,n}
%{\bf L}_1+{\bf F}_1{\bf D}_1+ {\bf D}_1^+{\bf E}_1+{\bf D}_1^+{\bf M}_1{\bf D}_1 
\end{array}\right), \quad  {\rm with} \quad 
{\bf L}_{1,n} = %{\bf L}_1+{\bf F}_1{\bf D}_1+ {\bf D}_1^+{\bf E}_1+{\bf D}_1^+{\bf M}_1{\bf D}_1 =
\left( \begin{array}{cc} 1 & \bar{\eta}  \\ 0 & 1 \end{array}\right),  
\end{eqnarray}
where ${\bf L}_{1,n}$ is the new longitudinal transfer matrix and 
$\bar{\eta}$ is a constant of the ring given by 
\begin{eqnarray}
\bar{\eta} &=& -\oint \frac{Dds}{\rho},   %
\end{eqnarray}
which is related to the usual momentum compaction factor $\alpha_0$ by 
$\bar{\eta}=-2\pi R \alpha_0$. We have shown that 
\begin{eqnarray}
\bar{\eta} &=& \eta_1 - \mathcal{H}_1 \sin 2\pi\nu_x,  
\end{eqnarray}
where $\nu_x$ is the betatron tune, $\eta_1$ is the $(1,2)$ element of ${\bf L}_{1}$ and 
the $\mathcal H$-function along with its associated phase parameter are defined by 
\begin{subequations} 
\begin{eqnarray}
\mathcal{H} &=& \frac1{\beta}[D^2 + (\alpha D + \beta D')^2], \\
\chi &=& \tan^{-1}\frac{ D}{\alpha D+ \beta D'},  
\end{eqnarray}
\end{subequations}
where $\alpha$, $\beta$ are Courant-Snyder parameters. 
We may define the fractional phase slippage factor on a section between point 1 and point 2 by 
\begin{eqnarray}
\bar{\eta}_{21} = -\int_{s_1}^{s_2} \frac{D ds}{\rho}. 
\end{eqnarray}
It has been shown that  
\begin{eqnarray}
\bar{\eta}_{21} = \eta_{21} - \sqrt{\mathcal{H}_1\mathcal{H}_2} \sin(\psi_{21}+\chi_1-\chi_2), 
\end{eqnarray}
where $\psi_{21}=\int_{s_1}^{s_2} ds/\beta$ is the betatron phase advance 
from point 1 to point 2.  

The new coordinates at any location after transformation are related 
to the original coordinates by 
\begin{subequations}
\begin{eqnarray}
{\bf x} &=& {\bf M} {\bf x}_n + \delta_n {\bf d},    \\
c\tau &=& c\tau_n + D x'_n-D' x_n, \\
\delta &=& \delta_n. 
\end{eqnarray}
\end{subequations}
We recognize ${\bf x}_n=(x_n, x'_n)^T$ are just the betatron coordinates. 
The momentum deviation coordinate is not changed by the transformation. But 
the longitudinal phase coordinate is different by  
\begin{eqnarray}
D x'_n-D' x_n  &=& -\sqrt{2 J_x \mathcal H}\cos(\psi - \chi),   
\end{eqnarray}
where $J_x$ is the horizontal betatron 
action variable and $\psi$ is the phase variable~\cite{SYLeeBook}. 

\section{Matrix formalism of synchrobetatron coupling \label{secMatForm}} 

When the rf cavity is turned on, its longitudinal transfer matrix deviates from the identity and 
is now
\begin{eqnarray}
{\bf L}_{\rm rf} &=& {\bf I} +{\bf W}, \qquad {\bf W} =  w \left( \begin{array}{cc} 0 & 0  \\ 1 & 0 \end{array}\right). 
\end{eqnarray}
Inserting it to Eq. (\ref{eqTGeneral}) we get the one-turn transfer matrix 
\begin{eqnarray}\label{eqTwithRF}
{\bf T} &=& {\bf T}^0 + w \tilde{\bf T}, \quad 
{\bf T}^0 = 
\left( \begin{array}{cc} {\bf M}^0_1 & {\bf E}^0_1  \\ 
{\bf F}^0_1   & {\bf L}^0_1 \end{array}\right), 
\end{eqnarray}
where the superscript $0$ denotes quantities when the rf cavity is off and 
\begin{eqnarray}\label{eqTwithRF2}
\tilde{\bf T} &=& 
\left( \begin{array}{cc} \tilde{\bf M}_1 & \tilde{\bf E}_1  \\ 
\tilde{\bf F}_1   &  \tilde{\bf L}_1 \end{array}\right) = \frac1w
\left( \begin{array}{cc} {\bf E}_{12}{\bf W}{\bf F}_{21} & {\bf E}_{12}{\bf W}{\bf L}_{21}  \\ 
{\bf L}_{12}{\bf W}{\bf F}_{21}   &  {\bf L}_{12}{\bf W}{\bf L}_{21} \end{array}\right). 
%\tilde{\bf M}_1 &=& \frac1w {\bf E}_{12}{\bf W}{\bf F}_{21} \\
%\tilde{\bf E}_1 &=& \frac1w {\bf E}_{12}{\bf W}{\bf L}_{21} \\
%\tilde{\bf F}_1 &=& \frac1w {\bf L}_{12}{\bf W}{\bf F}_{21} \\
%\tilde{\bf M}_1 &=& \frac1w {\bf L}_{12}{\bf W}{\bf L}_{21}. 
\end{eqnarray}

We may apply the procedure in Ref. \cite{SaganPRSTAB} directly to block diagonalize matrix $\bf T$ 
in Eq. (\ref{eqTwithRF}). However, it is easier to relate the elements in the transformation 
matrix to the well-known parameters such as dispersion functions and rf parameters in the following way. 
We first apply the transformation described in the previous section and then apply the 
procedure of Ref.~\cite{SaganPRSTAB} to decouple the new transfer matrix. 
After the first transformation, the off-diagonal blocks of the transfer 
matrix are small because the main dispersion effect is separated. Hence in the second transformation we 
can apply some approximations to derive explicit expressions for the transformation matrix and the final 
transfer matrix. 

After the first transformation, the transfer matrix is 
\begin{eqnarray}\label{eqTnwithRF}
{\bf T}_n = {\bf U}^{-1} {\bf T U} = {\bf T}^0_n + w\tilde{\bf T}_n, 
\end{eqnarray}
where
\begin{eqnarray}
{\bf T}^0_n &=& {\bf U}^{-1} {\bf T}^0{\bf U} = 
\left( \begin{array}{cc} {\bf M}_1^0 & {\bf 0} \\
{\bf 0} & {\bf L}^0_{1,n}
\end{array}\right) \quad {\rm and} \quad 
\tilde{\bf T}_n = {\bf U}^{-1} \tilde{\bf T}{\bf U} = 
\left( \begin{array}{cc} \tilde{\bf M}_n & \tilde{\bf E}_n \\
\tilde{\bf F}_n & \tilde{\bf L}_n   
\end{array}\right).    %\nonumber \\
%&=&
%\left( \begin{array}{cc} \tilde{\bf M}_1-{\bf D}_1\tilde{\bf F}_1-\tilde{\bf E}_1{\bf D}_1^++{\bf D}_1\tilde{\bf L}_1{\bf D}_1^+ & 
%\tilde{\bf E}_1+\tilde{\bf M}_1{\bf D}_1-{\bf D}_1\tilde{\bf L}_1-{\bf D}_1\tilde{\bf F}_1{\bf D}_1  \\
%\tilde{\bf F}_1+{\bf D}_1^+\tilde{\bf M}_1-\tilde{\bf L}_1{\bf D}_1^+-{\bf D}_1^+\tilde{\bf E}_1{\bf D}_1^+ & 
%\tilde{\bf L}_1+{\bf D}_1^+\tilde{\bf E}_1+\tilde{\bf F}_1{\bf D}_1+{\bf D}_1^+\tilde{\bf M}_1{\bf D}_1
%\end{array}\right).  
\end{eqnarray}
Following Sagan-Rubin~\cite{SaganPRSTAB}, we want to find the transformation matrix 
\begin{eqnarray}\label{eqVmat}
{\bf V} &=& 
\left( \begin{array}{cc} \gamma {\bf I} & {\bf C} \\
-{\bf C}^+ & \gamma {\bf I}
\end{array}\right)
\end{eqnarray}
to block diagonalize the matrix ${\bf T}_n$. 
According to the Sagan-Rubin procedure, we define  
\begin{eqnarray}
{\bf H}  &=& w (\tilde{\bf E}_n + \tilde{\bf F}^+_n) %{\bf E}_n + {\bf F}^+_n
\end{eqnarray}
and let 
\begin{eqnarray}\label{eqkappa}
\kappa = \frac{4 ||{\bf H}||}{{\rm Tr}\left[{\bf M}_n-{\bf L}_n  \right]^2}, 
\end{eqnarray}
where ${\bf M}_n$ and ${\bf L}_n$ are diagonal blocks of ${\bf T}_n$, 
the $\gamma$ parameter and ${\bf C}$ matrix are then 
\begin{eqnarray}\label{eqgammaC}
\gamma = \sqrt{\frac12+\frac12\sqrt{\frac1{1+\kappa}}},  \qquad
{\bf C} = -\frac{\bf H}{\gamma\sqrt{1+\kappa} {\rm Tr}\left[ {\bf M}_n-{\bf L}_n  \right] }. 
\end{eqnarray}
The elements of the ${\bf H}$ matrix have been calculated to be 
\begin{subequations}\label{eqHmat}
\begin{eqnarray}\label{eqHmat11}
H_{11} &=& -2w \sqrt{\beta_1 \mathcal{H}_2} \sin\pi \nu_x \cos(\pi \nu_x -\psi_{12}-\chi_2),   \\
H_{12} &=& - w\sqrt{\beta_1 \mathcal{H}_2} ( \bar{\eta} \sin(\psi_{12}+\chi_2) 
     - 2 \bar{\eta}_{12} \sin \pi \nu_x \cos(\pi\nu_x-\psi_{12}-\chi_2) ),   \\
H_{21} &=& -2w \sqrt{\frac{ \mathcal{H}_2}{\beta_1}} \sin \pi \nu_x [\sin(\pi \nu_x -\psi_{12}-\chi_2)-\alpha_1 
    \cos (\pi \nu_x -\psi_{12}-\chi_2)],   \\
H_{22} &=& w\sqrt{\frac{ \mathcal{H}_2}{\beta_1}}[ \bar{\eta}  (\alpha_1 \sin(\psi_{12}+\chi_2) - \cos(\psi_{12}+\chi_2))
   - 2 \bar{\eta}_{12} \sin \pi\nu_x  \nonumber \\
 &&   (\alpha_1 \cos(\pi\nu_x-\psi_{12}-\chi_2) - \sin(\pi\nu_x-\psi_{12}-\chi_2)) ]. 
 \label{eqHmat22}
\end{eqnarray}
\end{subequations}
It follows that 
\begin{eqnarray}\label{eqdetH}
||{\bf H}||=w^2\bar{\eta}\mathcal{H}_2 \sin 2\pi\nu_x. 
\end{eqnarray}
%Exact solution of the transformation matrix can be computed numerically with the above equations.  
%However, it is more illuminating to have an analytic solution, even with some approximations. 
Eqs. (\ref{eqVmat},\ref{eqkappa}-\ref{eqdetH}) constitute an analytic form of the decoupling 
transformation for the general case. In the following we simplify the expressions for the off-resonance 
cases in which ${\rm Tr}\left[ {\bf M}_n-{\bf L}_n  \right]$ is not close to zero.  
To this end we observe that $w$ is usually a small quantity (e.g., $w=0.008$~m$^{-1}$ for SPEAR3) and 
$\kappa$ is on the order of $w^2$ so that $\gamma\sqrt{1+\kappa}\approx 1+3\kappa/8=1+O(w^2)$. 
To first order of $w$ we have 
\begin{eqnarray}\label{eqCmat}
{\bf C} =-\frac{\bf H}{2(\cos 2\pi \hat{\nu}_x-\cos 2\pi\nu_s)},  
\end{eqnarray}
where we have used ${\rm Tr}[{\bf M}_n]=2\cos2\pi\hat{\nu}_x$ 
and ${\rm Tr}[{\bf L}_n]=2\cos2\pi\nu_s$ by definition. 
The parameter $\hat{\nu}_x$ is related to the unperturbed betatron tune $\nu_x$ by 
$\cos2\pi\hat{\nu}_x=\cos2\pi\nu_x+\frac12 w\mathcal{H}_2\sin2\pi\nu_x$ since 
it has been shown that 
%\begin{eqnarray}
${\rm Tr}[\tilde{\bf M}_n] = \mathcal{H}_2\sin 2\pi\nu_x$. 
%\end{eqnarray}
To guarantee the symplecticity of the new transfer matrix, we must have 
\begin{eqnarray}\label{eqapproxgamma}
\gamma &=& \sqrt{1-||{\bf C}||} 
= \sqrt{1-\frac{w^2 \bar{\eta}\mathcal{H}_2 \sin 2\pi\nu_x}{4(\cos 2\pi\hat{\nu}_x-\cos 2\pi\nu_s)^2}}. 
\end{eqnarray}

The transfer matrix for the decoupled coordinates may be written as 
\begin{eqnarray}
{\bf T}_d = {\bf V}^{-1} {\bf T}_n{\bf V} = 
\left( \begin{array}{cc} {\bf M}_d & {\bf 0} \\
{\bf 0} & {\bf L}_d
\end{array}\right),  
\end{eqnarray}
where ${\bf M}_d$ and ${\bf L}_d$ are given by \cite{SaganPRSTAB}
\begin{eqnarray}\label{eqMdLdSagan}
{\bf M}_d &=& \gamma^2 {\bf M}_n-\gamma ({\bf C} {\bf F}_n + {\bf E}_n {\bf C}^+) + {\bf C}{\bf L}_n{\bf C}^+,  \\
{\bf L}_d &=& \gamma^2 {\bf L}_n+\gamma ({\bf F}_n {\bf C}  +{\bf C}^+{\bf E}_n) + {\bf C}^+{\bf M}_n{\bf C}. 
\end{eqnarray}
For both of the above equations, the last three terms are on the order of $w^2$. To second order of 
$w$, we have 
\begin{eqnarray}\label{eqapproxMs}
{\bf M}_d &\approx& {\bf M}_1^0 + w\tilde{\bf M}_n + w^2 {\bf M}_d^{(2)}, \\ \label{eqapproxLs}
{\bf L}_d &\approx& {\bf L}_{1,n}^0 + w\tilde{\bf L}_n + w^2 {\bf L}_d^{(2)}. 
\end{eqnarray}
Explicit expressions have been developed for $\tilde{\bf M}_n$ and $\tilde{\bf L}_n$. 
The elements of $\tilde{\bf M}_n$ are 
\begin{subequations}
\begin{eqnarray}
\tilde{ M}_{n,11} &=& \mathcal{H}_2 \sin (\psi_{12}+\chi_2) (\cos(\psi_{21}-\chi_2)+\alpha_1 \sin(\psi_{21}-\chi_2)),  \\
\tilde{ M}_{n,12} &=& \mathcal{H}_2 \beta_1 \sin (\psi_{12}+\chi_2) \sin(\psi_{21}-\chi_2),  \\
\tilde{ M}_{n,21} &=& \frac{\mathcal{H}_2}{\beta_1} (\cos(\psi_{21}-\chi_2)+\alpha_1 \sin(\psi_{21}-\chi_2))
     (\cos(\psi_{12}+\chi_2)-\alpha_1 \sin(\psi_{12}+\chi_2)), \\
\tilde{ M}_{n,22} &=& \mathcal{H}_2 \sin (\psi_{21}-\chi_2) (\cos(\psi_{12}+\chi_2)-\alpha_1 \sin(\psi_{12}+\chi_2)). 
\end{eqnarray}
\end{subequations}
The additional terms $w\tilde{\bf M}_n$ amounts to changes to the Courant-Snyder parameters. 
For example, the change to $\beta_1$ to first order of $w$ is
\begin{eqnarray}
\Delta \beta_1 &=& \frac{w\mathcal{H}_2\beta_1}{2\sin2\pi\nu_x} (\cos (2\pi\nu_x-2\psi_{12}-2\chi_2)-\cos2\pi\nu_x).  
\end{eqnarray}
Also, knowing the traces 
%\begin{eqnarray}\label{eqTraceMs}
${\rm Tr}[\tilde{\bf M}_d^{(2)}] =-{\rm Tr}[\tilde{\bf L}_d^{(2)}]= 
-\bar{\eta}\mathcal{H}_2 \cot\pi\nu_x/2$, 
%-\frac{\bar{\eta}\mathcal{H}_2 \cos\pi\nu_x}{2 \sin \pi\nu_x}$
%\end{eqnarray}
we find the total betatron tune change to second order of $w$ 
\begin{eqnarray}\label{eqbetadetune}
\Delta \nu_x = -\frac{w \mathcal{H}_2}{4\pi} + \frac{w^2\bar{\eta}\mathcal{H}_2}{16\pi \sin^2\pi\nu_x}. 
\end{eqnarray}
From the expressions of ${\bf H}$ and $\tilde{\bf M}_n$, it is clear that when the rf cavity is located 
in a dispersion-free region and thus $\mathcal{H}_2=0$, there is no dynamic consequence from the 
coupling between horizontal and longitudinal motions. 

The matrix $\tilde{\bf L}_n$ is given by  
\begin{eqnarray}
\tilde{\bf L}_n &=& \left( \begin{array}{cc} \bar{\eta}_{12} & \bar{\eta}_{21}\bar{\eta}_{12} \\
1 & \bar{\eta}_{21}
\end{array}\right).     
\end{eqnarray}
The unperturbed longitudinal transfer matrix ${\bf L}_n={\bf L}^0_{1,n}+w \tilde{\bf L}_n $ 
describes the plain longitudinal motion. 
The corrections due to synchrobetatron coupling is 
on the next higher order of $w$. 
The matrix ${\bf L}_n$ can be Courant-Snyder parametrized so that 
\begin{eqnarray}
{\bf L}_n &=& %\left( \begin{array}{cc} 1+w\bar{\eta}_{12} & \bar{\eta}+w\bar{\eta}_{21}\bar{\eta}_{12} \\  w & 1+w\bar{\eta}_{21}
%\end{array}\right) =
\left( \begin{array}{cc} \cos \Phi_s+\alpha_s \sin\Phi_s & \beta_s\sin\Phi_s \\
-\gamma_s \sin\Phi_s & \cos \Phi_s - \alpha_s \sin\Phi_s
\end{array}\right),   
\end{eqnarray}
with $\beta_s\gamma_s=1+\alpha_s^2$ and $\Phi_s=-2\pi\nu_s$,  
where a negative sign is chosen to make $\beta_s$ always positive. 
The unperturbed synchrotron tune is given by
\begin{eqnarray}
\nu_s = \frac1{2\pi}\sin^{-1}\Big(\sqrt{-w\bar{\eta}(1+\frac14 w\bar{\eta})}\Big).  
%\approx \frac1{2\pi}\sqrt{-w\bar{\eta}}.  
\end{eqnarray}
The unperturbed longitudinal Courant-Snyder functions are %given by
\begin{eqnarray}
\alpha_s &=& -\frac12 \frac{w(\bar{\eta}_{12}-\bar{\eta}_{21})}{\sin2\pi\nu_s} , \quad 
\beta_s = -\frac{(\bar{\eta}+w \bar{\eta}_{12} \bar{\eta}_{21})}{\sin2\pi\nu_s },\quad
\gamma_s =  \frac{w}{\sin2\pi\nu_s},  
%\alpha_s &=& -\frac12 \sqrt{-\frac{w}{\bar{\eta}}} \frac{(\bar{\eta}_{12}-\bar{\eta}_{21})}{\sqrt{1+\frac14 w\bar{\eta}}} , \\
%\beta_s &=& -\frac1{\sqrt{-w\bar{\eta}}} \frac{(\bar{\eta}+w \bar{\eta}_{12} \bar{\eta}_{21})}{\sqrt{1+\frac14 w\bar{\eta}}},\\
%\gamma_s &=& \sqrt{-\frac{w}{\bar{\eta}}} \frac1{\sqrt{1+\frac14 w\bar{\eta}}}.  
\end{eqnarray}
which are equivalent forms of those found in Ref.~\cite{BNashThesis}. 
It is noted that $\gamma_s$ is a positive constant and $\beta_s$ is positive but slightly varies 
around the ring. 
Without coupling, the longitudinal coordinates $(c\tau_n, \delta_n)$ are related to these parameters, 
the longitudinal action variable $J_s$ and phase 
variable $\psi_s$ by 
\begin{eqnarray} 
c\tau_n = \sqrt{2J_s \beta_s} \cos \psi_s, \quad
\delta_n = -\sqrt{\frac{2J_s}{\beta_s}} (\alpha_s \cos \psi_s+\sin\psi_s), 
\end{eqnarray}
from which we obtain relations between the rms bunch length, rms momentum 
spread and the longitudinal emittance $\epsilon_s$ 
\begin{eqnarray}
\sigma^2_{c\tau_n} = \beta_s \epsilon_s, \qquad \sigma^2_{\delta_n} = \gamma_s \epsilon_s. 
\end{eqnarray}
Note that the bunch length varies with location because 
particles with different momentum deviation experience different longitudinal phase slippage. 
The bunch is longest at the rf cavity and can be shortened by a maximum of $\pi^2\nu^2_s/2$ part of the original 
length at half way across the ring from the cavity. 
For fast ramping synchrotrons, it can be as large as 5\%
, assuming $\nu_s=0.1$. The synchrobetatron coupling should 
slightly change the synchrotron tune and the longitudinal Courant-Snyder functions given above. 

The decoupled coordinates ${\bf X}_d = ({\bf x}^T_d, {\bf l}^T_d)^T$ are related to the original 
coordinates ${\bf X}$ by 
\begin{eqnarray}\label{eqTransfromXXs}
{\bf X} &=& {\bf UV}{\bf X}_d = 
\left( \begin{array}{cc} \gamma {\bf I}-{\bf D}_1{\bf C}^+ & {\bf C}+\gamma{\bf D}_1 \\
-{\bf C}^+ - \gamma{\bf D}_1^+ & \gamma {\bf I}-{\bf D}_1^+{\bf C} 
\end{array}\right)  
\left( \begin{array}{c} {\bf x}_d \\ {\bf l}_d
\end{array}\right).   
\end{eqnarray}
The betatron coordinates $({\bf x}^T_n, {\bf l}^T_n)^T$ are related to the normal modes 
by the transformation matrix ${\bf V}$. 
Since all four elements of the ${\bf C}$ matrix in ${\bf V}$ are nontrivial in general, 
the horizontal betatron coordinate $x_n$ depends on the longitudinal phase, 
as pointed out in Ref.~\cite{GuoXL}. This is a natural consequence of the synchrobetatron 
coupling. In fact, the longitudinal coordinates $c\tau_n$, $\delta_n$ also depend on 
the horizontal betatron coordinates. In terms of the betatron coordinates 
${\bf X}_n$, the phase space volume occupied by the beam tilts across the horizontal 
and longitudinal subspace. Particles flow in and out between the two subspaces. However, the total 
phase space volume is preserved. The reason for Ref.~\cite{GuoXL} to suggest that the 
total phase space volume is not preserved is because it didn't fully consider the coupling effects on 
the horizontal betatron motion. 

With Eq. (\ref{eqTransfromXXs}) we can decompose the 4-dimension motion to the normal modes 
for any given initial condition. 
It also allows us to derive the effects of synchrobetatron coupling on the beam sizes $\sigma_x$ 
and $\sigma_{c\tau}$. 
Following the single mode analysis of Ref. \cite{SaganPRSTAB}, we first consider the case when 
only the betatron mode, or mode $a$,  is excited so that  
\begin{eqnarray}
\left( \begin{array}{c} x \\ x'
\end{array}\right) &=& (\gamma {\bf I}-{\bf D}_1{\bf C}^+) 
\left( \begin{array}{c} x_a \\ x'_a
\end{array}\right), 
\qquad 
\left( \begin{array}{c} c\tau \\ \delta
\end{array}\right) = -({\bf C}^+ + \gamma{\bf D}_1^+ )  
\left( \begin{array}{c} x_a \\ x'_a 
\end{array}\right),  
\end{eqnarray}
where $(x_a, x'_a)$ are the betatron normal mode coordinates given by 
\begin{eqnarray}
x_a &=&   \sqrt{2J_a \beta_a} \cos \psi_a, \qquad
x'_a =  -\sqrt{\frac{2J_a}{\beta_a}}(\alpha_a \cos\psi_a+\sin\psi_a). 
\end{eqnarray}
It follows that  
\begin{subequations}
\begin{eqnarray}
\frac{x}{\sqrt{2J_a}} &=& [\gamma\sqrt{\beta_a} + D (C_{21}\sqrt{\beta_a}+C_{11}\frac{\alpha_a}{\sqrt{\beta_a}})] 
           \cos\psi_a + \frac{D C_{11}}{\sqrt{\beta_a}} \sin \psi_a, \\
\frac{c\tau}{\sqrt{2J_a}} &=& -[\gamma (\beta_a D'+\alpha_a D)+(\beta_a C_{22}+\alpha_a C_{12})]\frac1{\sqrt{\beta_a}} \cos\psi_a 
	- \frac{(C_{12}+\gamma D)}{\sqrt{\beta_a}}\sin\psi_a. %\\
\end{eqnarray}
\end{subequations}
The bunch width and length can be derived from the above equations by integrating over the bunch distribution. 
Here we consider only the off-resonance cases and assume $\cos{\nu_s}\approx 1$ so that 
${\bf C}\approx{\bf H}/4\sin^2\pi\nu_x$. We get 
\begin{eqnarray}\label{eqsigmaxa}
\sigma^2_{x,a} 
&=& \beta_a \epsilon_a - \epsilon_a \frac{wD\sqrt{\beta_1\mathcal{H}_2}}{\sin\pi\nu_x}
	\sin(\pi\nu_x-\psi_{12}-\chi_2) + \epsilon_a \frac{w^2\mathcal{H}_2}{4\sin^2\pi\nu_x}  
    (D^2- \frac12 \beta_1\bar{\eta} \cot\pi\nu_x), 
\end{eqnarray}
where $\epsilon_a$ is the emittance of the horizontal betatron normal mode. 
We see that the term on the order of $w$ varies around the ring. The leading non-varying 
correction term is on the order of $w^2$. Since the $O(w^2)$ terms 
are very small, we will drop them in the following for brevity. 
Similarly for the bunch length we obtain  
\begin{eqnarray}\label{eqsigmactaua}
\sigma^2_{c\tau,a}  &=&  \epsilon_a \mathcal{H}^{a}_1 - \epsilon_a \frac{w\sqrt{\mathcal{H}_1 \mathcal{H}_2}}{2\sin^2\pi\nu_x}
[\bar{\eta}\cos(\psi_{12}+\chi_2-\chi_1)-2\bar{\eta}_{12} \sin\pi\nu_x \sin(\pi\nu_x-\psi_{12}-\chi_2   % \nonumber \\
  +\chi_1)], \nonumber \\ 	%+ \epsilon_a \frac{w^2\mathcal{H}_2}{16\sin^4\pi\nu_x} 
%	(\bar{\eta}^2-4\bar{\eta}_{12} \bar{\eta}_{21}\sin^2\pi\nu_x - \bar{\eta}\mathcal{H}_1 \sin2\pi\nu_x),  	
\end{eqnarray}
where $\mathcal{H}^{a}_1$ is the $\mathcal{H}$-function at point 1 
evaluated with the perturbed Courant-Snyder functions. 

The same analysis can be applied to the synchrotron mode, or mode $b$, 
to obtain its contributions to bunch width and length, which are given by 
\begin{eqnarray}\label{eqsigmaxb}
\sigma^2_{x,b} &=& \epsilon_b\gamma_b D^2 + 
\epsilon_b \gamma_b\frac{ w\bar{\eta} D\sqrt{\beta_1 \mathcal{H}_2}}{2\sin^2\pi\nu_x}
	  \cos\pi\nu_x \sin(\pi\nu_x-\psi_{12}-\chi_2), \\
\label{eqsigmactaub}	
\sigma^2_{c\tau,b} &=& \epsilon_b\beta_b - \epsilon_b \gamma_b \frac{\bar{\eta}\sqrt{\mathcal{H}_1\mathcal{H}_2}}{\sin\pi\nu_x} 
	\cos(\pi\nu_x-\psi_{12}-\chi_2+\chi_1).  
\end{eqnarray}
 Note that $\gamma_b$ is an $O(w)$ factor. 
The momentum spread is given by
\begin{eqnarray} %\label{eqsigmadeltaa}
\sigma^2_{\delta,a}  &=&  \epsilon_a \frac{w^2 \mathcal{H}_2}{4 \sin^2 \pi\nu_x}, \qquad
\label{eqsigmadeltab}  
\sigma^2_{\delta,b} =  \epsilon_b\gamma_b - \epsilon_b\gamma_b\frac{w^2 \bar{\eta}\mathcal{H}_2 \sin2\pi\nu_x}{16\sin^4\pi\nu_x}. 
\end{eqnarray}
So to first order of $w$ we have $\sigma^2_{\delta} =  \epsilon_b\gamma_b$. 

The contributions to $\sigma^2$ from mode $a$ and $b$ simply add up~\cite{SaganPRSTAB}, i.e.
\begin{eqnarray}
\sigma^2 = \sigma^2_a + \sigma^2_b, 
\end{eqnarray}
which applies to all four coordinates. 
Eqs. (\ref{eqsigmaxa}-\ref{eqsigmadeltab})  show that because of synchrobetatron coupling,  
the longitudinal motion affects the transverse beam size and and 
the transverse motion affects the longitudinal beam size. 
The leading correction term is usually a small term on the order of $w$ and varies with 
the horizontal betatron phase advance around the ring. 
It is noted that we have recovered Shoji's result in Ref. \cite{ShojiBunLen} by the first 
term in Eq. (\ref{eqsigmactaua}) which is independent of $w$ and 
indicates that the bunch length varies from location to location according to the 
local $\mathcal{H}$-function. 
This is a consequence of the uneven distribution of the 
path length effect. 

\section{Closed-orbit change due to energy gain and loss \label{secCOD}}

The finite energy gain at the rf cavity and the radiation energy loss around the ring 
are kicks to the momentum deviation coordinate. These kicks are transfered downstream 
and affect all other coordinates. Therefore the closed orbit of the beam is changed. 
To study this effect in the matrix formalism, we extend the coordinate vector to 
${\bf X}^e = (x, x', c\tau, \delta, 1)^T$~\cite{AChao79}. The corresponding transfer matrix is then 
5$\times$5. The fifth element is included to describe the kicks received by the particle, 
namely the equation  
\begin{eqnarray}
{\bf X}_2 &=& {\bf T}_{21} {\bf X}_1 + {\bf g}_{21} 
\end{eqnarray}
will now be written as 
\begin{eqnarray}
{\bf X}^e_2 &=& {\bf T}^e_{21} {\bf X}^e_1,  \qquad 
 {\bf T}^e_{21} = \left( \begin{array}{cc} {\bf T}_{21} & {\bf g}_{21} \\
{\bf 0} & 1 \end{array}\right) 
\end{eqnarray}
where ${\bf g}_{21}$ is a 4-vector which represents the kick-induced shifts of phase space 
coordinates across the accelerator section from point 1 to point 2. 
The closed orbit ${\bf X}_c$ is given by the fixed point ${\bf X}_c^e=({\bf X}^T_c,1)^T$ of 
the extended one-turn transfer matrix ${\bf T}^e$ 
~\cite{AChao79}, i.e., ${\bf T}^e{\bf X}_c^e={\bf X}_c^e$, 
which yields
\begin{eqnarray} \label{eqInv4Xc}
{\bf X}_c = ( {\bf I} - {\bf T} )^{-1} {\bf g}, 
\end{eqnarray} 
where the ${\bf g}$ vector contains the first four elements of the fifth column of ${\bf T}^e$ and 
represents the coordinate shifts after one turn when a particle starts 
from a point with initial coordinates of all zeros. We will derive analytic forms for 
${\bf g}$ and $({\bf I} - {\bf T} )^{-1}$ below. 

For an rf cavity, there is an energy kick 
\begin{eqnarray}
{\bf g}_{\rm rf} = (0,0,0,\epsilon)^T, \qquad \epsilon = \frac{\Delta E}E, 
\end{eqnarray}
where $\epsilon$ denotes the sudden change of momentum deviation at the rf cavity. 
For a dipole magnet, the energy kick due to radiation energy loss will propagate to the 
other coordinates and cause finite changes to them. For example, the 
pure sector dipole with bending radius $\rho$ and bending angle $\theta_L$ has 
\begin{eqnarray}
{\bf g}_{\rm dipole} = (-\frac{\epsilon}{2\pi})
\left( \begin{array}{c} \rho(\theta_L-\sin\theta_L) \\1-\cos \theta_L \\ 
\rho (1-\cos \theta_L-\frac12\theta^2_L) \\
\theta_L
\end{array}\right), 
\end{eqnarray}
where $-\epsilon \theta_L/2\pi$ is the momentum deviation change on this dipole magnet. 
Here we have assumed $\epsilon$  is the same as the momentum deviation gained at the rf cavity. 
It is seen that the changes of $x$, $x'$ and $c\tau$ are on the order of $\theta^3_L$, 
$\theta^2_L$ and $\theta^4_L$, respectively. Hence when $\theta_L$ is small, we may neglect 
these changes for a single dipole magnet. The $\bf g$ vectors are zeros for other accelerator 
components which don't cause energy gains or losses. 

The one turn extended transfer matrix can be computed as usual. 
For an arbitrary point 1 , we have 
%\begin{eqnarray}
${\bf T}^e_1 = {\bf T}^e_{12} {\bf T}^e_{\rm rf} {\bf T}^e_{21}$, 
%\end{eqnarray}
from which we obtain 
\begin{eqnarray} \label{eqg1sum}
{\bf g}_1 = ({\bf T}_{12} {\bf g}_{21} + {\bf g}_{12}) + {\bf T}_{12} {\bf g}_{\rm rf} + 
{\bf T}_{12} ({\bf T}_{\rm rf}-{\bf I}){\bf g}_{21}. 
\end{eqnarray}
Simple calculations show that 
\begin{eqnarray} 
{\bf T}_{12} {\bf g}_{\rm rf} &=& \left( \begin{array}{cc} {\bf M}_{12} & {\bf E}_{12}  \\ 
{\bf F}_{12}   &  {\bf L}_{12}  \end{array}\right) 
\left( \begin{array}{c} 0 \\ 0 \\0 \\ \epsilon
 \end{array}\right) = \epsilon 
\left( \begin{array}{c} {\bf d}_1 - {\bf M}_{12}{\bf d}_2 \\  
\left( \begin{array}{c} \eta_{12} \\ 1
 \end{array}\right) 
 \end{array}\right).  
\end{eqnarray}
The ${\bf T}_{12} {\bf g}_{21} + {\bf g}_{12}$ term has nothing to do with the 
rf cavity. So it does not depend on the location of point 2. In fact it represents   
the changes of coordinates in one turn for a particle with initial 
coordinate $(0,0,0,0)^T$ at point 1 when the rf cavity is turned off. 
Since energy losses occur in dipoles, we get a summation over all dipoles 
in the ring
\begin{eqnarray} 
{\bf T}_{12} {\bf g}_{21} + {\bf g}_{12} &=& 
\sum_{i=1}^{N_d} \left( \begin{array}{cc} {\bf M}_{1i} & {\bf E}_{1i}  \\ 
{\bf F}_{1i}   &  {\bf L}_{1i}  \end{array}\right) 
\left( \begin{array}{c} 0 \\ 0 \\0 \\ -\epsilon_i
 \end{array}\right) = - 
\left( \begin{array}{c} \sum_i {\bf e}_{1i}\epsilon_i \\  
\left( \begin{array}{c} \sum_i \eta_{1i} \epsilon_i \\ \sum_i \epsilon_i
 \end{array}\right) 
 \end{array}\right).  
\end{eqnarray}
Noting that ${\bf e}_{1i} = {\bf d}_1 - {\bf M}_{1i}{\bf d}_i$ and, assuming 
all bending radius are equal so that $\epsilon_i = \epsilon \Delta s_i/2\pi\rho$, 
the summations can be turned to integrals to obtain 
\begin{eqnarray} 
{\bf T}_{12} {\bf g}_{21} + {\bf g}_{12} &=& (-\epsilon)  
\left( \begin{array}{c} {\bf d}_1 - \oint_1 {\bf M}(s_1|s) {\bf d}(s) \frac{ds}{2\pi\rho} \\  
\left( \begin{array}{c} \oint_1 \eta(s_1|s) \frac{ds}{2\pi\rho} \\ 1
 \end{array}\right) 
 \end{array}\right),   
\end{eqnarray}
where we have used $\sum_i \epsilon_i=\epsilon$ and the integral $\oint_1$ starts 
from point 1, completes one revolution and ends at point 1. 
We may simplify the expressions by 
defining new functions 
\begin{subequations} \label{eqSCKxi}
\begin{eqnarray}  \label{eqSCKxi1}
\mathcal{S}(s) &=& \int_s^{s+C} \sqrt{\mathcal{H}(s')} \sin(\psi_{ss'}+\chi_{s'})\frac{ds'}{2\pi\rho},  \\
\mathcal{C}(s) &=& \int_s^{s+C} \sqrt{\mathcal{H}(s')} \cos(\psi_{ss'}+\chi_{s'})\frac{ds'}{2\pi\rho},  \\
\mathcal{K}(s) &=& { \mathcal{S}^2(s) + \mathcal{C}^2(s)} , \qquad 
\xi(s) = \tan^{-1} \frac{\mathcal{S}(s)}{\mathcal{C}(s)},   \label{eqSCKxi3}
\end{eqnarray}
\end{subequations}
where $\psi_{ss'}$ is the betatron phase advance from point $s'$ to point $s$ and 
$C$ is the ring circumference. 
Then we can write down 
\begin{eqnarray} 
\oint_1 {\bf M}(s_1|s) {\bf d}(s) \frac{ds}{2\pi\rho} = 
 \left( \begin{array}{cc} \sqrt{\beta_1} & 0 \\ -\frac{\alpha_1}{\sqrt{\beta_1}} & \frac1{\sqrt{\beta_1}}
 \end{array}\right)  \left( \begin{array}{c} \mathcal{S}_1 \\ \mathcal{C}_1
 \end{array}\right) 
\end{eqnarray}
and 
\begin{eqnarray} 
\oint_1 \eta(s_1|s) \frac{ds}{2\pi\rho} &=& \oint_1 \bar{\eta}(s_1|s)\frac{ds}{2\pi\rho}+
\sqrt{\mathcal{H}_1\mathcal{K}_1} \sin(\xi_1-\chi_1). 
%\sqrt{\mathcal{H}_1}(\mathcal{S}_1\cos\chi_1-\mathcal{C}_1 \sin \chi_1). 
\end{eqnarray}
It can be shown that 
\begin{eqnarray} 
\oint_1 \bar{\eta}(s_1|s)\frac{ds}{2\pi\rho} = \frac12 \bar{\eta}. 
\end{eqnarray} 
Since the first three elements of the ${\bf g}$ vector have terms on the order of $O(\epsilon)$ 
and the contributions of the third term  of Eq. (\ref{eqg1sum})  to these elements are on 
the order of $O(w\epsilon)$, we will simply drop these contributions. However, 
the fourth element of this term is the leading term and is not negligible. 
It is easy to show that 
\begin{eqnarray} 
g^4_1 &=& w g^3_{21} = -w\epsilon \int_{s_1}^{s_2} {\eta}(s_2|s)\frac{ds}{2\pi\rho}  \nonumber \\
&=&   -w\epsilon \int_{s_1}^{s_2} \bar{\eta}(s_2|s)\frac{ds}{2\pi\rho} - 
%w\epsilon \sqrt{\mathcal{H}_2} \big(\mathcal{S}_{21}\cos\chi_2 
% -\mathcal{C}_{21}\sin\chi_2 \big),  
w \epsilon \sqrt{\mathcal{H}_2 \mathcal{K}_{21}} \sin(\xi_{21}-\chi_2) 
 \label{eq4thg1} 
\end{eqnarray} 
with the definition of 
\begin{subequations}
\begin{eqnarray}
\mathcal{S}_{21} &=& \int_{s_1}^{s_2}\sqrt{\mathcal{H}(s')}\sin(\psi_{s_2s'}+\chi_{s'})\frac{ds}{2\pi\rho}, \\
\mathcal{C}_{21} &=& \int_{s_1}^{s_2}\sqrt{\mathcal{H}(s')}\cos(\psi_{s_2s'}+\chi_{s'})\frac{ds}{2\pi\rho}, \\ 
{\mathcal K}_{21} &=& \mathcal{S}_{21}^2+ \mathcal{C}_{21}^2,\qquad 
\xi_{21} = \tan^{-1} \frac{\mathcal{S}_{21}}{\mathcal{C}_{21}}. 
\end{eqnarray}
\end{subequations}
If we write down the vector ${\bf g}_1$ in a form ${\bf g}_1 = ({\bf g}^T_x, {\bf g}^T_l)^T$ with 
2-component vector ${\bf g}_x$ and ${\bf g}_l$ representing the horizontal and longitudinal 
displacement, respectively, we have 
%\begin{subequations}
%\begin{eqnarray} \label{eqgxgl}
%{\bf g}_x/\epsilon &=& \oint_1 {\bf M}(s_1|s) {\bf d}(s) \frac{ds}{2\pi\rho} - {\bf M}_{12} {\bf d}_2,  \\
%{\bf g}_l/\epsilon &=& \left( \begin{array}{c} \eta_{12} - \oint_1 {\eta}(s_1|s)\frac{ds}{2\pi\rho} \\ 
%-w \int_{s_1}^{s_2} {\eta}(s_2|s)\frac{ds}{2\pi\rho} 
% \end{array}\right),  
%\end{eqnarray} 
%\end{subequations}
%which can also be written as 
\begin{subequations}\label{eqg1vec}
\begin{eqnarray} \label{eqgxvec}
{\bf g}_x/\epsilon &=& \left( \begin{array}{cc} \sqrt{\beta_1} & 0 \\ -\frac{\alpha_1}{\sqrt{\beta_1}} & \frac1{\sqrt{\beta_1}}
 \end{array}\right)  
 \left( \begin{array}{c} \mathcal{S}_1-\sqrt{{\mathcal H}_2}\sin(\psi_{12}+\chi_2) 
 \\ \mathcal{C}_1-\sqrt{{\mathcal H}_2}\cos(\psi_{12}+\chi_2) 
 \end{array}\right),  \\   \label{eqgl1}
g_l^1/\epsilon &=& \bar{\eta}_{12}-\frac12\bar{\eta} + \sqrt{{\mathcal H}_1 {\mathcal H}_2}
\sin(\psi_{12}+\chi_2-\chi_1) - \sqrt{{\mathcal H}_1 {\mathcal K}_1}
\sin(\xi_1-\chi_1) , \\   \label{eqgl2}
g_l^2/\epsilon &=& -w \frac{\bar{\eta}^2_{21}}{2\bar{\eta}} -  
w \sqrt{\mathcal{H}_2 \mathcal{K}_{21}} \sin(\xi_{21}-\chi_2) , 
\end{eqnarray} 
\end{subequations}
where $g_l^1$, $g_l^2$ are the two elements of ${\bf g}_l=(g_l^1,g_l^2)^T$ and in obtaining 
the first term of Eq. (\ref{eqgl2}) we have assumed the dipoles are distributed around the ring uniformly. 

Radiation damping changes the transfer matrix ${\bf T}$ by a correction term 
on the order of $\epsilon$ so that it is no more strictly symplectic. 
However, since $\epsilon$ is usually small and the correction to the closed orbit due to 
this effect is on the order of $\epsilon^2$, we will not consider the radiation damping effect. 
To work out the matrix inversion for $( {\bf T} - {\bf I} )^{-1}$, 
we will make use of 
\begin{eqnarray} \label{eqTmIandTnmI}
({\bf T} - {\bf I})^{-1} &=& {\bf U} ({\bf T}_n - {\bf I})^{-1} {\bf U}^{-1}, 
\end{eqnarray} 
since the off-diagonal blocks of ${\bf T}_n$ are on the order of $O(w)$. 
%The inverse of a general 4$\times$4 matrix can be expressed in terms of its 2$\times$2 blocks. 
With matrix ${\bf T}_n$ as found in Eq. (\ref{eqTnwithRF}), we have shown that 
\begin{eqnarray} 
%\left( \begin{array}{cc} {\bf \mathcal A} & {\bf  \mathcal B} \\ {\bf \mathcal  C} & {\bf \mathcal  D}
% \end{array}\right)^{-1} = 
({\bf T}_n - {\bf I})^{-1} &=& \left( \begin{array}{cc} {\bf a} & {\bf b} \\ {\bf c} & {\bf d}
 \end{array}\right)  
\end{eqnarray} 
with 
\begin{subequations} 
\begin{eqnarray} 
{\bf a} &= &({\bf M}^0_1 - {\bf I})^{-1}, \\
{\bf b} &= &-({\bf M}^0_1 - {\bf I})^{-1} \left( \begin{array}{cc} 0 & {\tilde E}^{11}_n \\ 0 & {\tilde E}^{21}_n
 \end{array}\right),    \\
{\bf c} &= &-\left( \begin{array}{cc} {\tilde F}^{21}_n & {\tilde F}^{22}_n \\ 0 & 0
 \end{array}\right) ({\bf M}^0_1 - {\bf I})^{-1}, \\
{\bf d} &= &({\bf L}_n - {\bf I})^{-1},  
\end{eqnarray} 
\end{subequations}
where we have dropped all terms on the order of $O(w)$ or higher and we have 
\begin{eqnarray} 
({\bf L}_n - {\bf I})^{-1} &=& \frac1{\bar{\eta}}  
 \left( \begin{array}{cc} -\bar{\eta}_{21} & \frac{\bar{\eta}}w + \bar{\eta}_{21} \bar{\eta}_{12} \\ 
 1 & -\bar{\eta}_{12}
 \end{array}\right).   
\end{eqnarray}
We then proceed to obtain the blocks of $({\bf T-I})^{-1}$ using Eq. (\ref{eqTmIandTnmI}) and finally we 
get the closed orbit with Eq. (\ref{eqInv4Xc}). The results are given by  
\begin{subequations}
\begin{eqnarray} \label{eqAnaXc1}
x_c &=&  -\frac{\epsilon \sqrt{\beta_1 \mathcal{H}_2}}{2\sin\pi\nu_x} \cos(\pi\nu_x-\psi_{12}-\chi_2)  
   + \frac{\epsilon\sqrt{\beta_1 \mathcal{K}_1}}{2\sin\pi\nu_x} \cos(\pi\nu_x-\xi_1)  
    + \epsilon D_1 (\frac{1}2-\frac{\bar{\eta}_{12}}{\bar{\eta}}),  \\
x'_c &=& \frac{\epsilon}{2\sin\pi\nu_x}\sqrt{\frac{\mathcal{H}_2}{\beta_1}}
\big(\alpha_1 \cos(\pi\nu_x-\psi_{12}-\chi_2)-
	\sin(\pi\nu_x-\psi_{12}-\chi_2) \big)-  \nonumber \\
	&&\frac{\epsilon}{2\sin\pi\nu_x}\sqrt{\frac{\mathcal{K}_1}{\beta_1}}
	\big(\alpha_1\cos(\pi\nu_x-\xi_1)-\sin(\pi\nu_x-\xi_1)\big)
	+\epsilon D'_1 (\frac{1}2-\frac{\bar{\eta}_{12}}{\bar{\eta}}), \\  \label{eqAnaAc3}
c\tau_c &=& \epsilon \frac{\bar{\eta}_{21} \bar{\eta}_{12}}{2\bar{\eta}}
  +\frac{\epsilon}{{2\sin\pi\nu_x}} \big[\sqrt{\mathcal{H}_1\mathcal{H}_2}\cos(\pi\nu_x-\psi_{12}-\chi_2+\chi_1)    
   +\sqrt{\mathcal{K}_1\mathcal{H}_2}\cos(\pi\nu_x-  \psi_{12}  -\chi_2 \nonumber \\ 
   &&+\xi_1)
   -\sqrt{\mathcal{K}_1\mathcal{H}_1}\cos(\pi\nu_x+\chi_1-\xi_1) 
   - \mathcal{H}_2 \cos\pi\nu_x  \Big] + \epsilon \sqrt{\mathcal{H}_2 \mathcal{K}_{21}} \sin(\xi_{21}-\chi_2),   \\ \label{eqAnaXc4}
\delta_c &=&  \epsilon (\frac{1}2-\frac{\bar{\eta}_{12}}{\bar{\eta}}).  
\end{eqnarray}
\end{subequations}
Note again that we have dropped all terms on the order of $w$ or higher 
so that the results are valid only for off-resonance cases. 
Obviously we have 
recovered Eqs. (23-24) of Ref. \cite{GuoXL} as the first term in Eq. (\ref{eqAnaXc1}). 
The second term comes from the propagation of radiation energy losses. The third term 
comes from the energy variation around the ring. It is worth noting that 
the $c\tau_c$ orbit is zero at the rf cavity. 

So far we have considered only one rf cavity in the ring. However, the same analysis can be easily 
applied to more cavities. In fact, if we neglect the interaction between the rf cavities, which corresponds to 
higher order terms of the $w_i$ parameters, the one-turn transfer matrix is 
\begin{eqnarray}\label{eqT1multirf}
{\bf T}_1 &=& {\bf T}^0_1+\sum_i {\bf T}_{1i}{\bf W}_i{\bf T}_{i1}, 
\end{eqnarray}
where the summation is over all cavities. Eq. (\ref{eqT1multirf}) means the total effect of all  
cavities is the linear superposition of the single cavity effects. 
It is then straightforward to modify the results of the one-cavity case for 
multiple-cavity cases. 
 
\section{Simulation \label{secSim}}

In this section we will verify the theory developed in the previous sections by 
comparing the results to simulations with the 
accelerator modeling code AT \cite{AccelTool}. We use the machine model of the 
SPEAR Booster for the calculation. The SPEAR Booster is chosen because it has appreciable dispersion 
functions at the rf cavity. The model consists of 20 periods of FODO lattice over a 
circumference of $2\pi R=133.8$~m. 
The bending radius is $\rho=11.82$~m for all dipoles. The extraction energy is $E=3$~GeV. 
The rf frequency is $f_{\rm rf}=358.533$~MHz and the harmonic number is $h=160$. At extraction, 
the rf gap voltage is $V_{\rm rf}=0.8$~MV and the one-turn radiation energy loss is $U_0=0.60$~MeV. 
In the simulation, we will consider it as a storage ring running at its extraction energy. 
The betatron tunes are $\nu_x=6.16$ and $\nu_y=4.23$ for the model. 
At the rf cavity, the Courant-Snyder functions are $\alpha_2=-0.72$ and $\beta_2=2.09$~m 
and the dispersion functions are $D_2=0.27$~m and $D'_2=0.06$. The horizontal 
chromaticity is $C_x=-7.9$. 

We first present a numerical example to verify the transformation Eq. (\ref{eqTransfromXXs}). 
The injection point of the SPEAR Booster is located in the ring opposite to the 
rf cavity where 
we have $\alpha_1=0.02$, $\beta_1=1.54$~m, $D_1=0.24$~m and $D'_1=0.01$ and 
the betatron phase advance from the rf cavity to this observation point is $\psi_{12}=0.16$~rad 
modulo $2\pi$.  The synchronous phase is set to $\phi_s=\pi$ and hence $w=0.0020$~m$^{-1}$. 
The one-turn transfer matrix  
for betatron coordinates ${\bf X}_n$ at this location is 
\begin{eqnarray}
{\bf T}_n &=& \left(
\begin{array}{rrrr}
 \text{0.550532 } & \text{1.306266 } & -\text{0.000447} &
   \text{0.001018} \\
 -\text{0.547728} & \text{0.516813} & \text{0.000129} &
   -\text{0.000294} \\
 \text{0.000396 } & -\text{0.000910} & \text{0.995433} &
   -\text{4.548267} \\
 -\text{0.000174} & \text{0.000399 } & \text{0.002004} &
   \text{0.995433}
\end{array}
\right). 
\end{eqnarray}
The off-diagonal blocks represents the synchrobetatron coupling effects. We then apply the 
second transformation ${\bf V}$ as determined by 
Eqs. (\ref{eqVmat}, \ref{eqHmat}, \ref{eqCmat}-\ref{eqapproxgamma}) to obtain the 
final transfer matrix 
\begin{eqnarray}
{\bf T}_d &=& \left(
\begin{array}{rrrr}
 \text{0.550533} & \text{1.306265} &  -4.6\times10^{-10}  & 1.05\times10^{-9}  \\
 -\text{0.547728} & \text{0.516813} & 1.3\times10^{-10} & -3.0\times10^{-10}  \\
 4.1\times10^{-10}  & -9.4\times10^{-10}  &
   \text{0.995432 } & -\text{4.548266 } \\
 -1.8\times10^{-10}  & 4.1\times10^{-10}  &
   \text{0.002004} & \text{0.995432 }
\end{array}
\right). 
\end{eqnarray}
Elements of the off-diagonal blocks are reduced to the order of magnitude of $10^{-10}$ 
from the original value of $10^{-4}$, indicating the high precision of the approximations we have 
made.  

\begin{figure}[hbp]
   \begin{center}
      \includegraphics[width=4.0in]{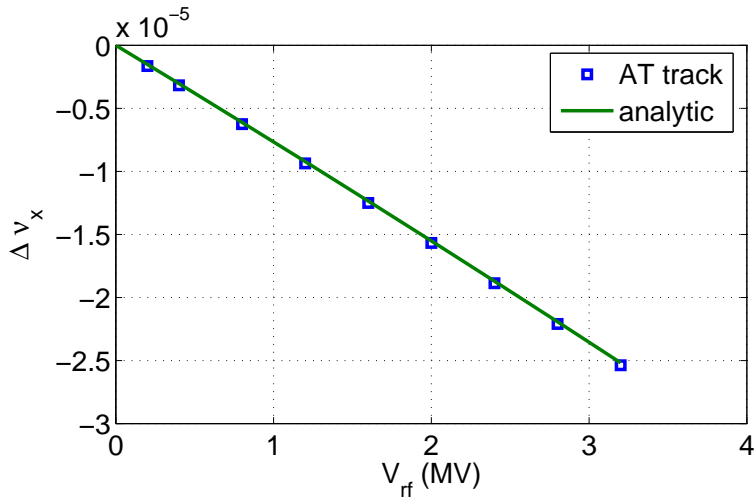}
      \caption{ The synchrobetatron coupling induced betatron detuning $\Delta \nu_x$ obtained 
      by AT tracking is compared to analytic calculation with Eq. (\ref{eqbetadetune}). 
      The rf gap voltage is varied from 0.2~MV to 3.2~MV.     }
      \label{figcmpnuxTrackSB}
   \end{center}
\end{figure}

The betatron detuning due to synchrobetatron coupling (Eq. (\ref{eqbetadetune})) is checked against AT 
by comparing the betatron tune from AT tracking to the analytic calculation with 
various rf voltage. The initial condition for AT tracking is 
$( 1\times10^{-5}, 0, 0, 0,  1\times10^{-6}, 0)^T$. Small deviations of $x$ and $\delta$ 
are chosen to avoid significant nonlinear detuning effects. 
The result is shown in Figure \ref{figcmpnuxTrackSB}. 

\begin{figure}[hbp]
   \begin{center}
      \includegraphics[width=4.0in]{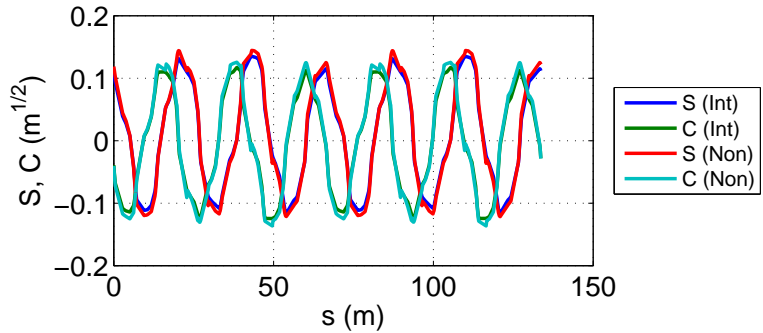}
      \caption{ (Color online) Function ${\mathcal S}$ and ${\mathcal C}$ as defined in Eq. (\ref{eqSCKxi}) 
      for the SPEAR Booster at 
      3 GeV. The curves labeled ``Int'' are from numerical integration with the Trapezoidal rule by 
      cutting each dipole into 100 slices. The ``Non'' curves are obtained by calculating the coordinate 
      shifts at the exit of each dipole, transporting to the observation point and summing up contributions of 
      all dipoles.    }
      \label{figSB_ShCh}
   \end{center}
\end{figure}

\begin{figure}[hbp]
   \begin{center}
      \includegraphics[width=3.0in]{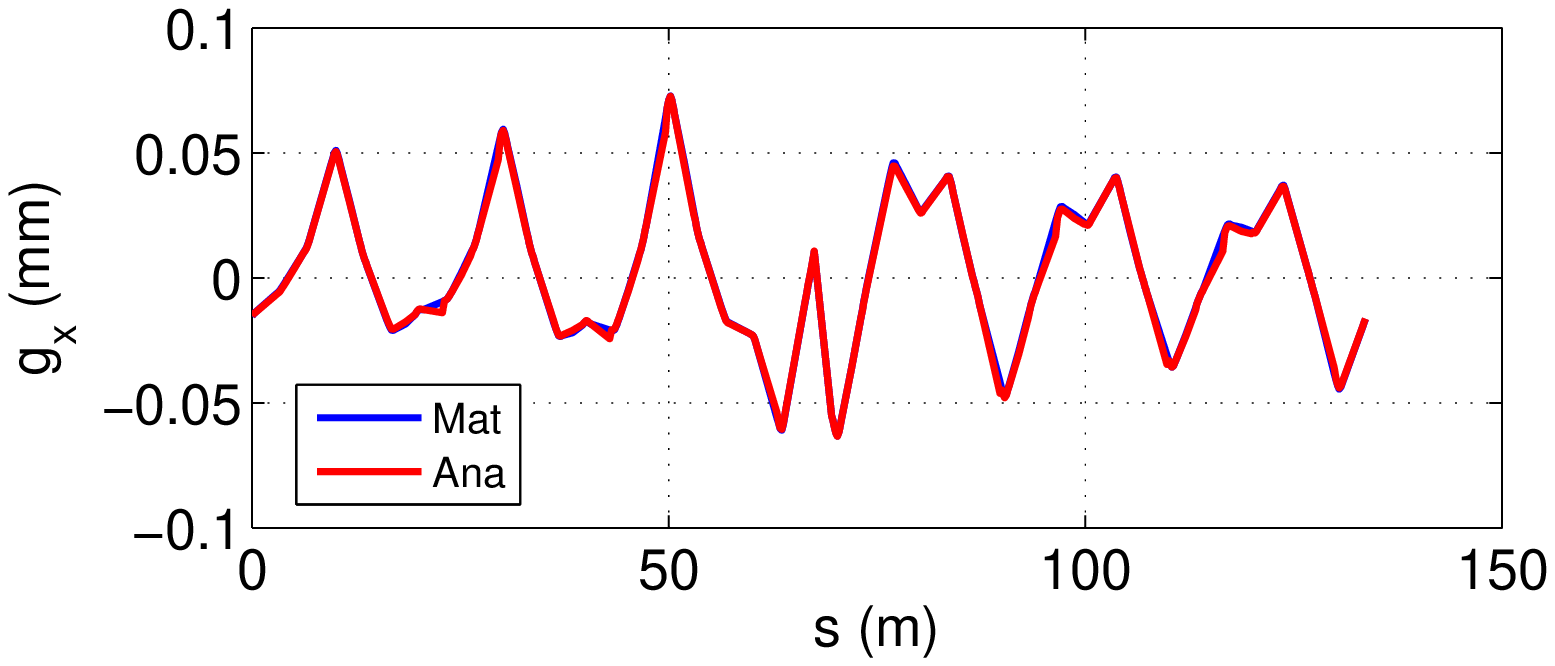}
      \includegraphics[width=3.0in]{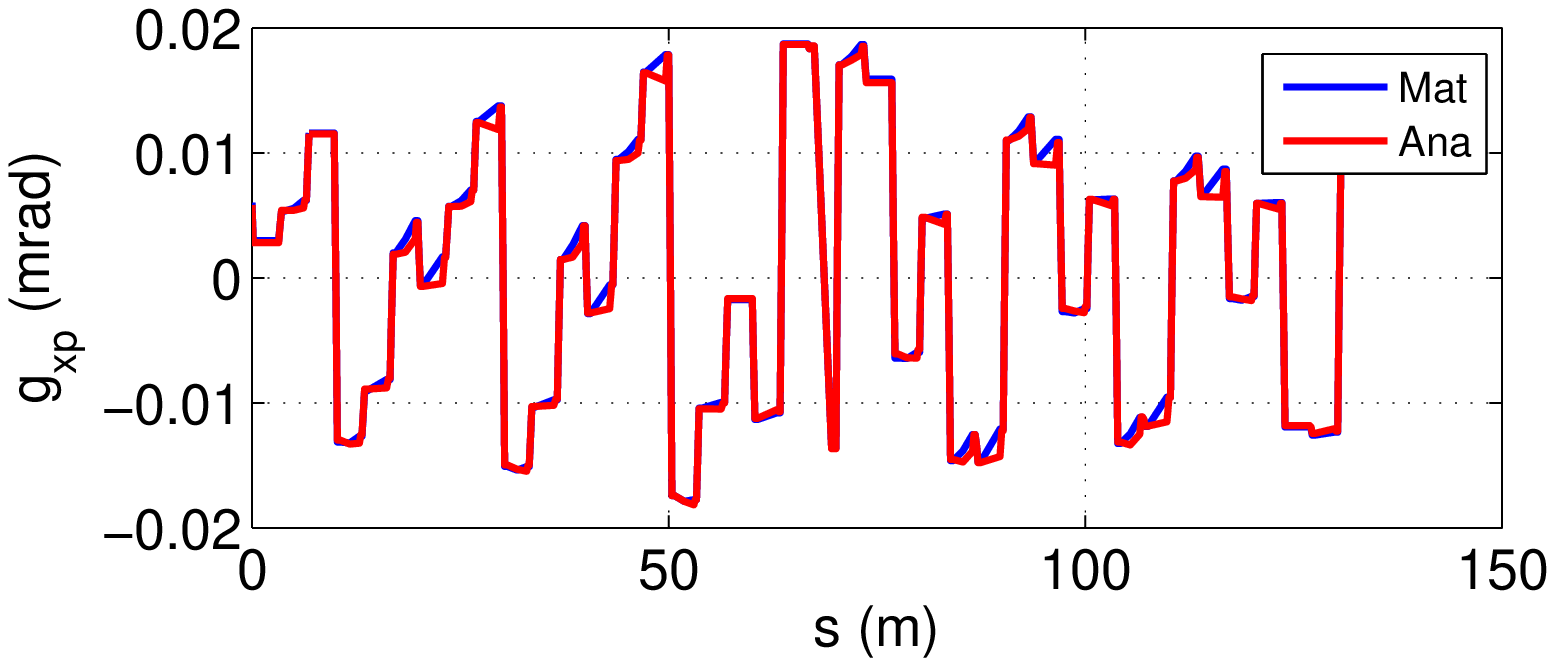}
      \includegraphics[width=3.0in]{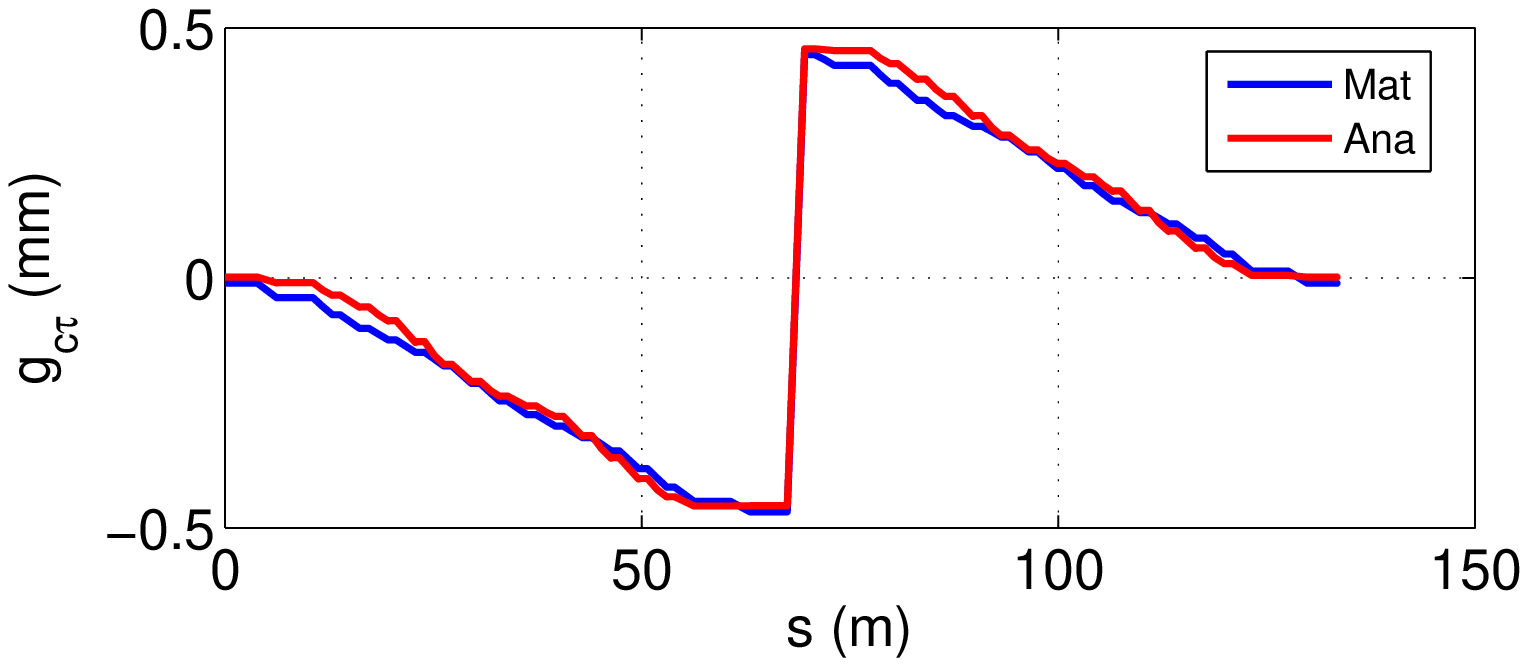}
      \includegraphics[width=3.0in]{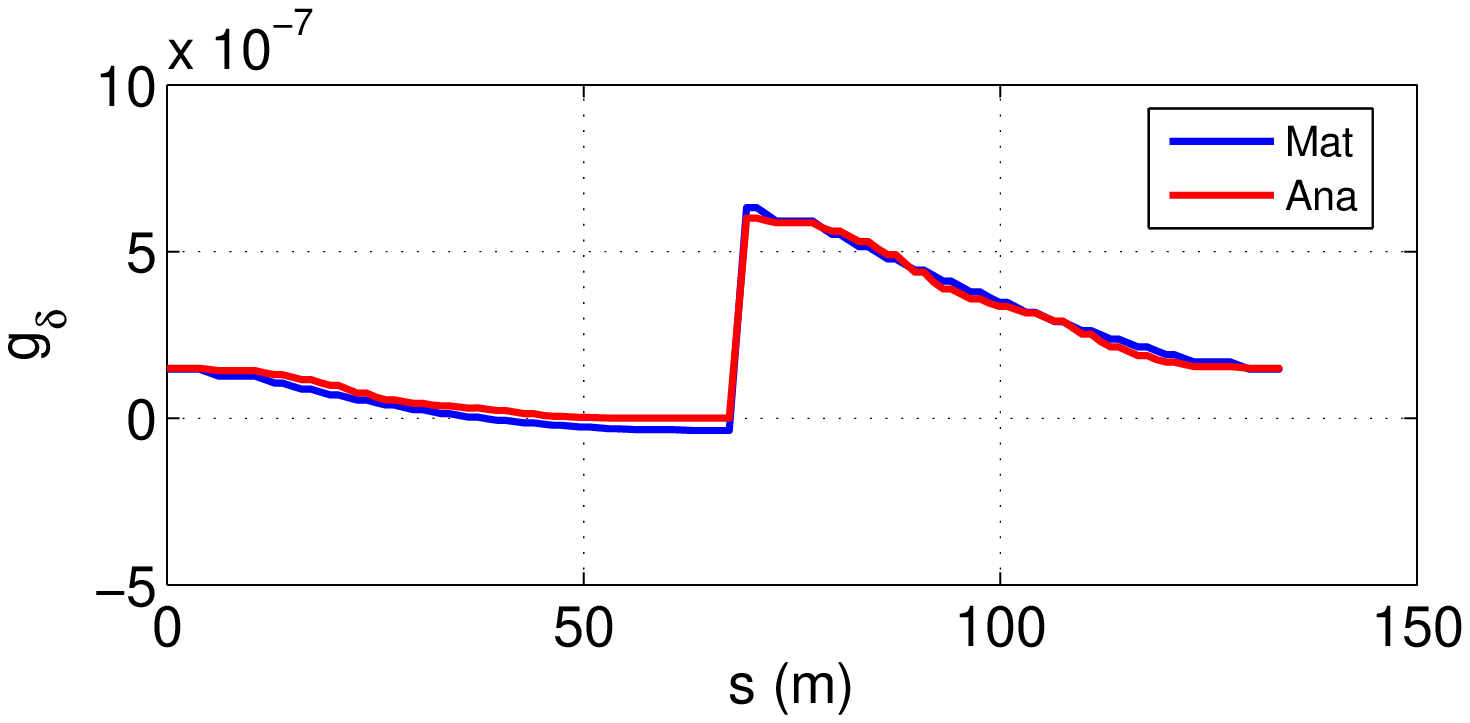}
      \caption{ (Color online) The one-turn coordinate shifts due to radiation energy losses and 
      finite energy gains at the rf cavity for 
      the SPEAR Booster at 3 GeV. The four plots are $x$, $x'$, $c\tau$ and $\delta$ shifts, respectively.     }
      \label{figSB_gvec}
   \end{center}
\end{figure}

If radiation is turned on in simulation, the finite energy 
gain at the rf cavity will cause changes to the closed orbit as studied in the previous section. 
To calculate the closed orbit change, we need to calculate the functions 
${\mathcal S}$, ${\mathcal C}$, ${\mathcal K}$ and $\xi$ defined 
in Eq. (\ref{eqSCKxi}). It is found that numerical integration through a dipole
does not make much difference from simply transporting the coordinate shifts at the dipole exit 
to the observation point. In Figure \ref{figSB_ShCh} we show functions 
${\mathcal S}$ and ${\mathcal C}$ for the SPEAR Booster ring 
calculated with both methods. 
The one-turn coordinate shifts, or the ${\bf g}$ vector is plotted in Figure \ref{figSB_gvec}. 
There are two curves in each plot, one is from Eq. (\ref{eqg1vec}), the 
other is from direct matrix multiplication  with the transfer 
matrix of each accelerator element given by AT and the coordinate shifts of each element obtained by 
tracking through it with zero initial coordinates in AT.  

\begin{figure}[hbp]
   \begin{center}
      \includegraphics[width=3.0in]{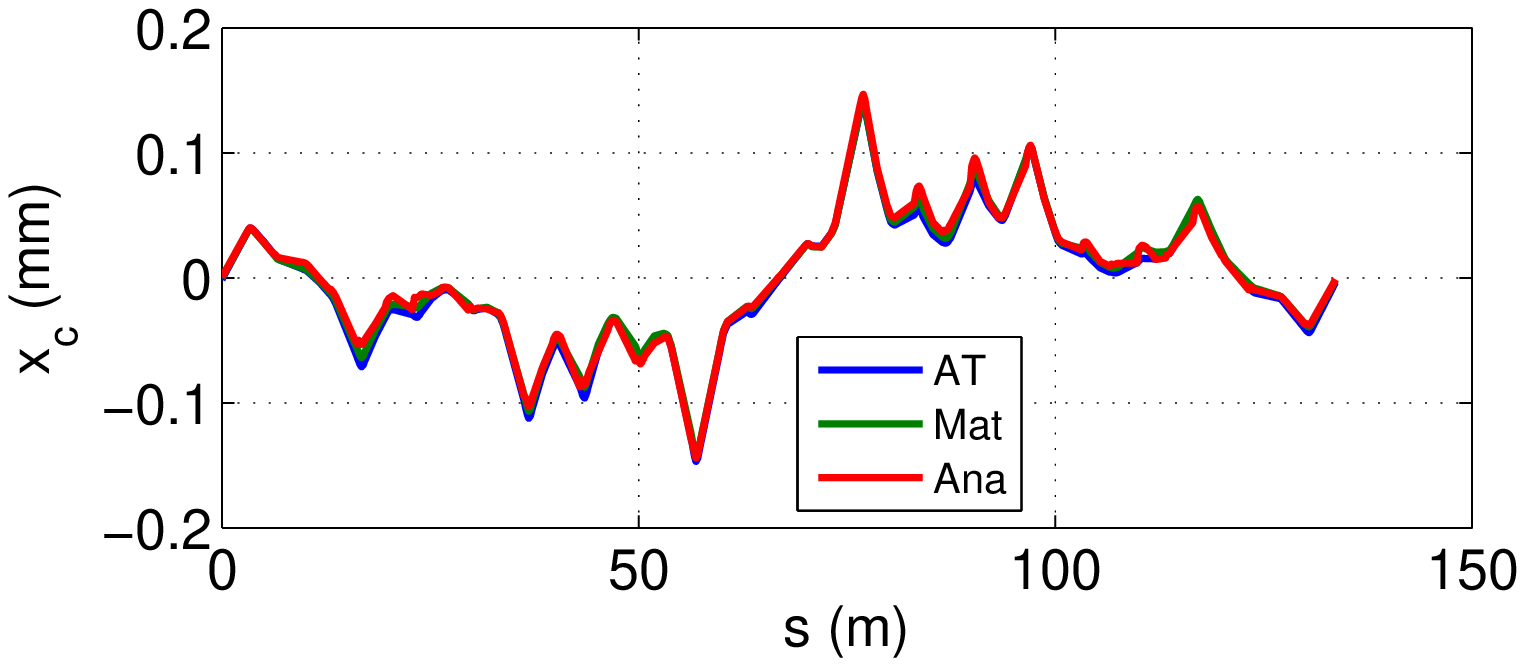}
      \includegraphics[width=3.0in]{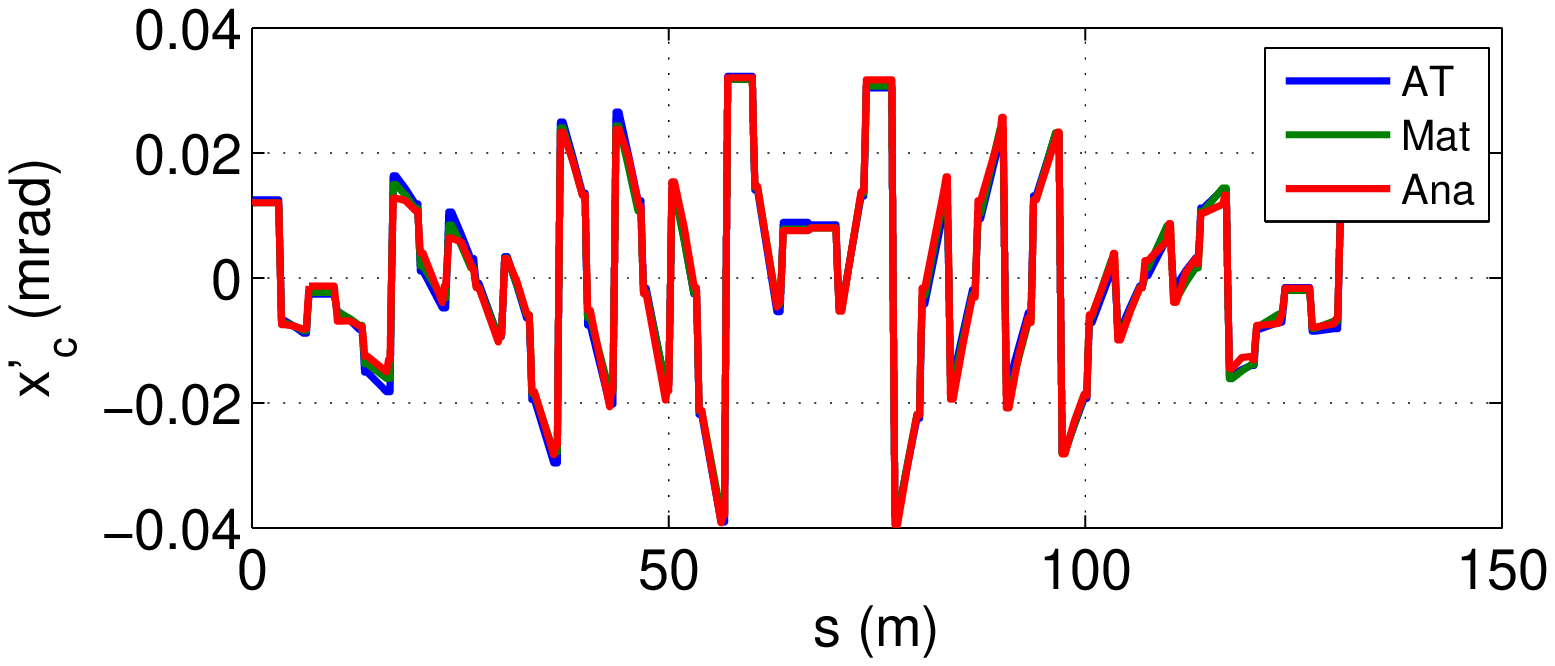}
      \includegraphics[width=3.0in]{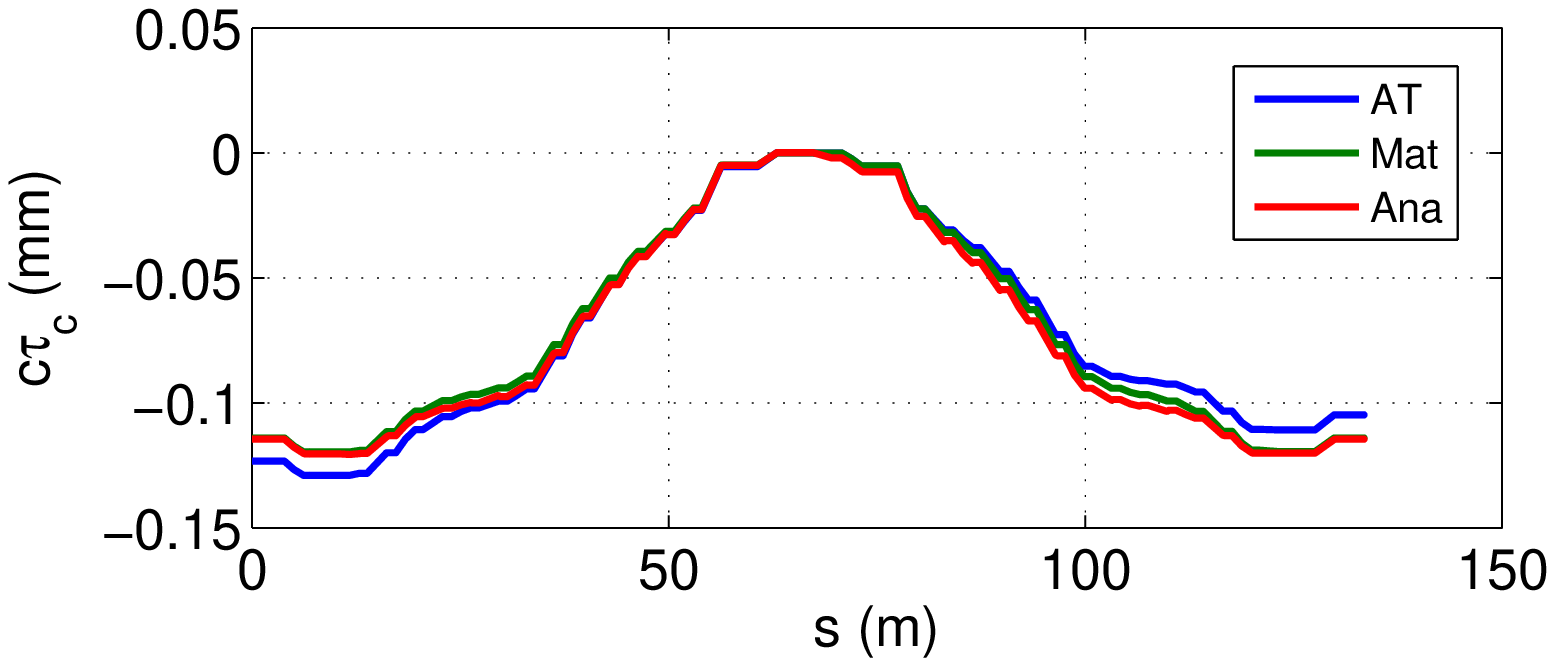}
      \includegraphics[width=3.0in]{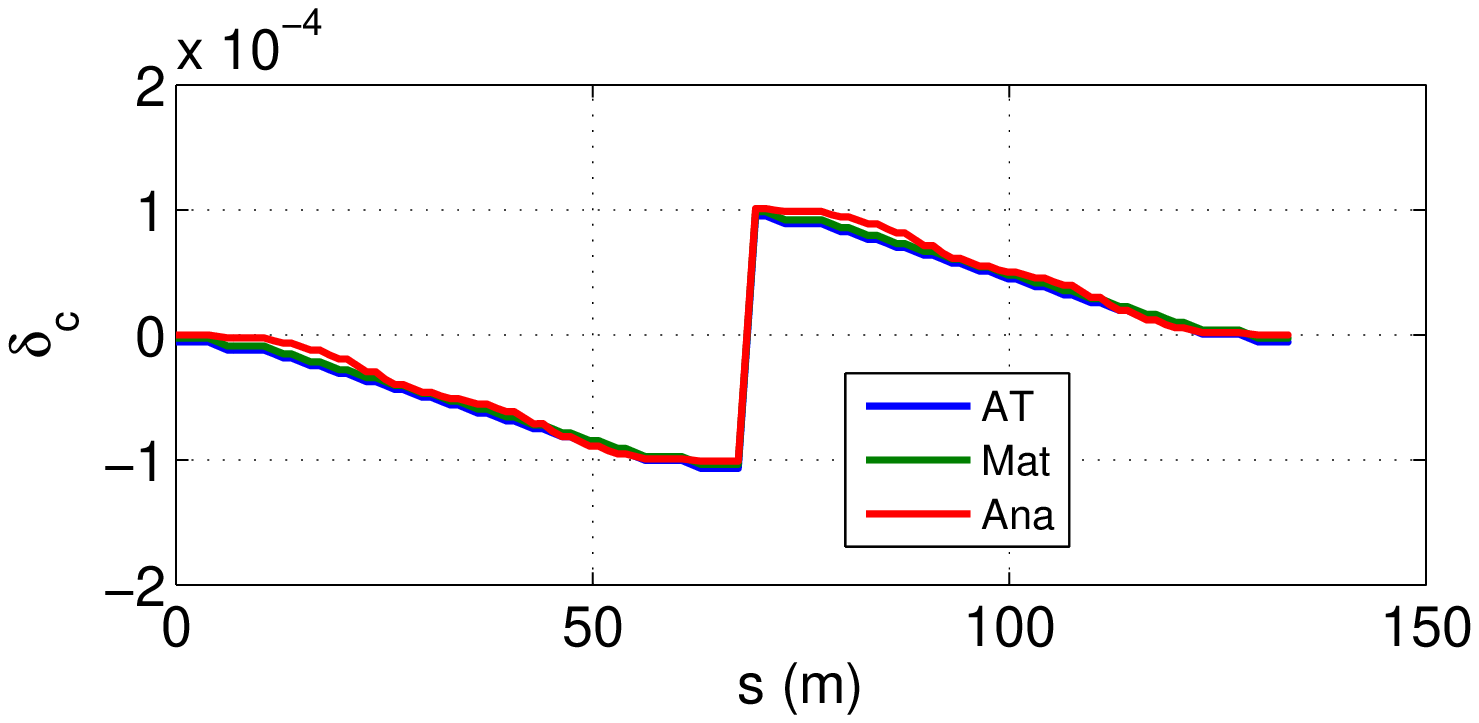}
      \caption{ (Color online) The SBC induced closed orbit changes calculated in three different ways 
      for the SPEAR Booster at 3 GeV. The changes to $x_c$, $x'_c$, $c\tau_c$ and $\delta_c$ are shown 
      in the four plots. 
      The three curves are from AT (``AT''), matrix inversion (``Mat'') and the analytic solution (``Ana''), respectively.      }
      \label{figSB_Xc}
   \end{center}
\end{figure}

We then compare the closed orbit changes obtained with three different ways: (1) AT (using the function {\em findorbit6}), 
(2) direct matrix calculation with Eq. (\ref{eqInv4Xc}) and 
(3) the analytic solution in Eqs. (\ref{eqAnaXc1}-\ref{eqAnaXc4}). 
The results are shown in Figure \ref{figSB_Xc}. Good agreement between the three curves are seen, verifying 
the analysis in the last section. 
The ${\mathcal S}$ and ${\mathcal C}$ functions are calculated with simple summations in 
the analytic approach. It is worth pointing out that all the three terms in Eq. (\ref{eqAnaXc1}) are important in determining
the closed orbit change for $x_c$. 

\begin{figure}[hbp]
   \begin{center}
      \includegraphics[width=3.0in]{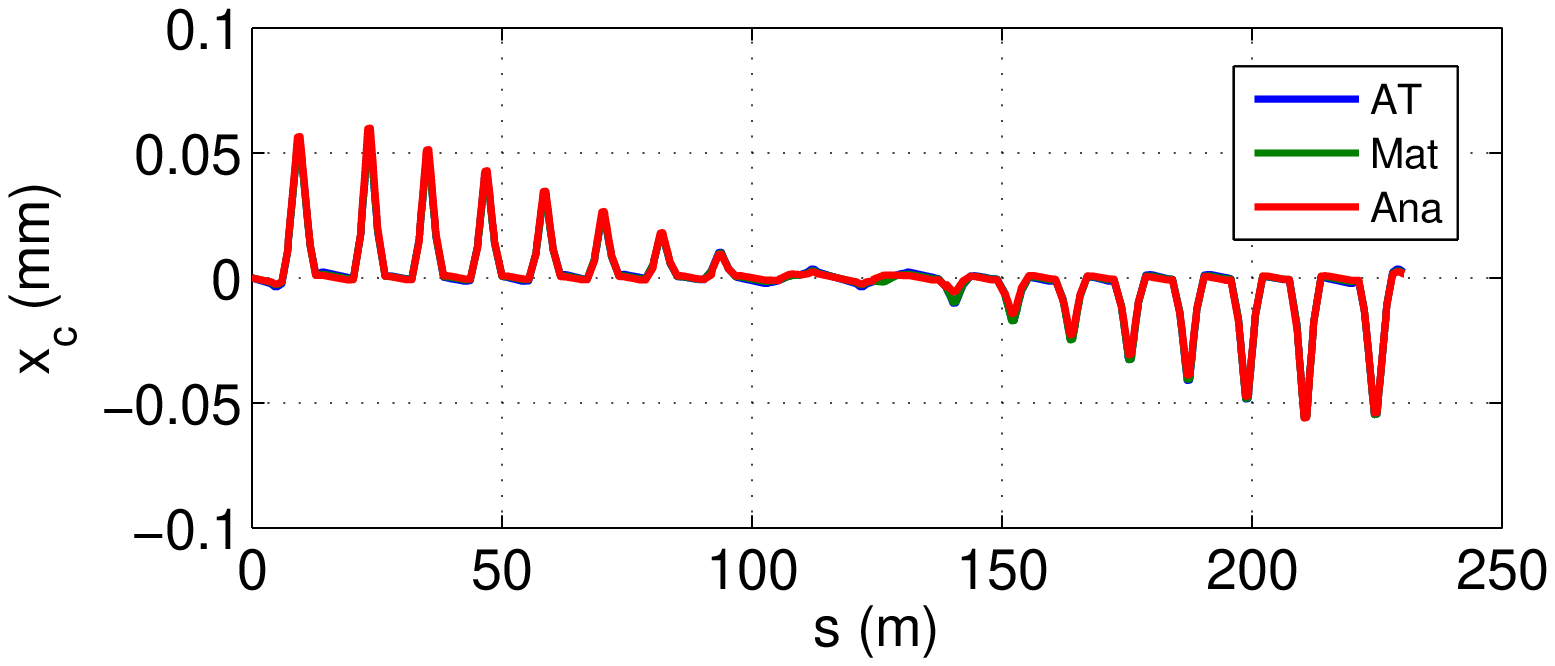}
      \includegraphics[width=3.0in]{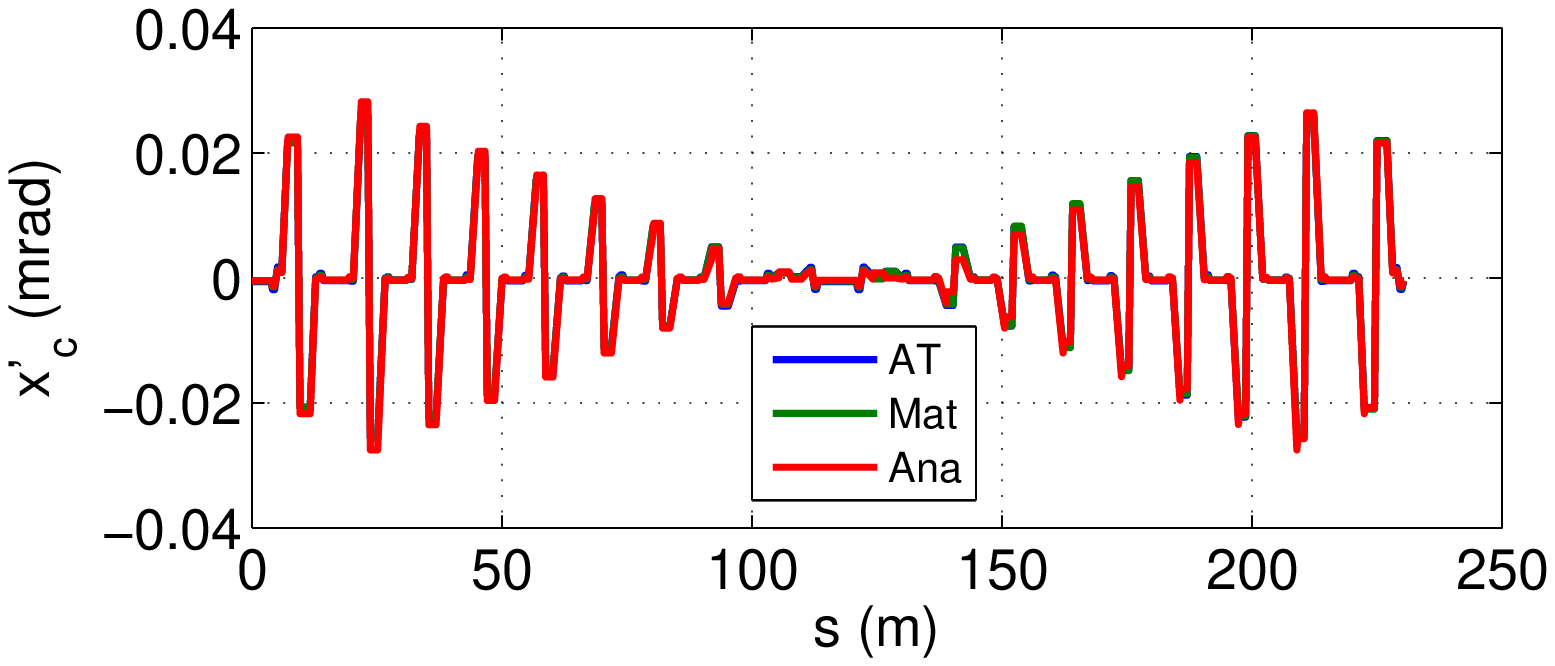}
      \includegraphics[width=3.0in]{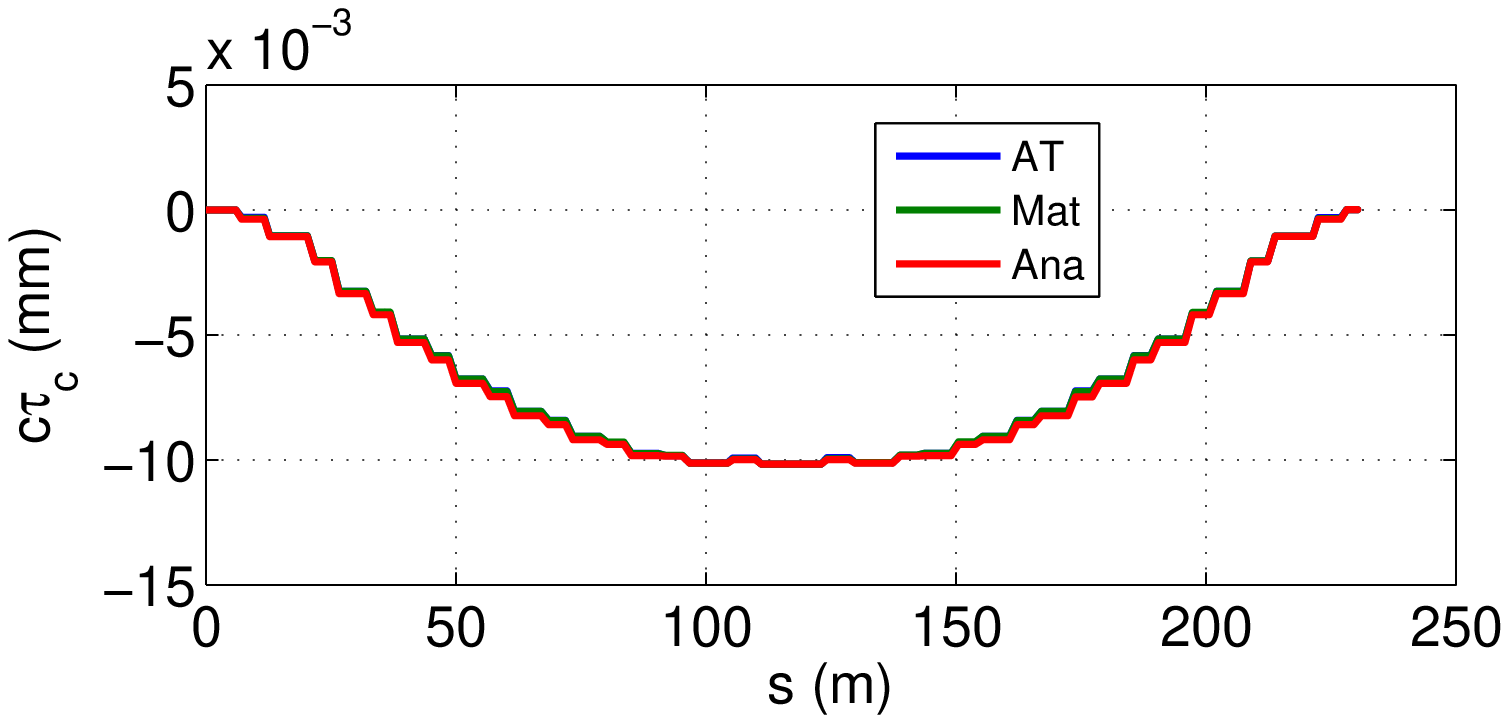}
      \includegraphics[width=3.0in]{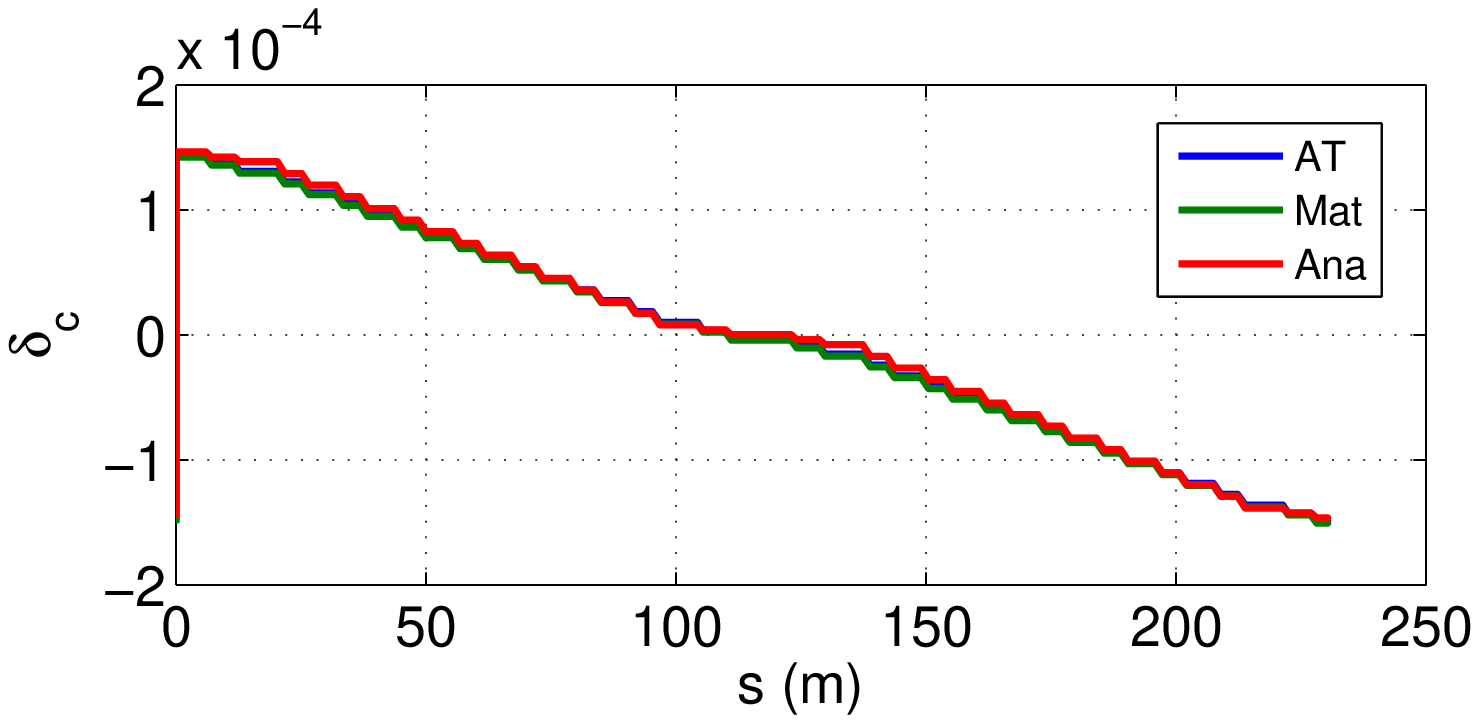}
      \caption{ (Color online) The SBC induced closed orbit changes calculated in three different ways 
      for the SPEAR3 ring. The changes to $x_c$, $x'_c$, $c\tau_c$ and $\delta_c$ are shown. 
      The three curves are from AT (``AT''), matrix inversion (``Mat'') and the analytic solution (``Ana''), respectively.      }
      \label{figSB_Xc_sp3}
   \end{center}
\end{figure}

The same closed orbit calculation is done for the SPEAR3 ring. We only show 
the closed orbit changes in Figure \ref{figSB_Xc_sp3}. 
Note dispersion at the rf cavity for the SPEAR3 ring is zero. So the terms 
in Eqs. (\ref{eqAnaXc1}-\ref{eqAnaXc4}) involving ${\mathcal H}_2$ have no contribution.  
 
\section{Conclusions \label{secDiscu}}

In this study we fully analyzed the linear synchrobetatron coupling by block diagonalizing 
the $4\times4$ horizontal-longitudinal transfer matrix. 
We found the transformation between the usual $(x, x', c\tau, \delta)$ coordinates and the 
normal modes and the transfer matrix for the normal modes in analytic forms in terms of 
the Courant-Snyder functions, dispersion functions and the rf voltage slope parameter. 
This enabled us to predict the 4-dimensional motion of a particle knowing only the initial condition 
and those common parameters. 
The effects of synchrobetatron coupling on the horizontal betatron motion, including changes 
to the Courant-Snyder functions and the betatron tune are also presented. 
We then studied the beam width and the bunch length under synchrobetatron coupling. We readily 
recovered Shoji's result of dispersion-dependent bunch lengthening~\cite{ShojiBunLen}. 
We found that the beam width and the bunch length slightly oscillate around the ring with the 
betatron phase advance measured from the rf cavity. We pointed out that
the bunch length varies around the ring due to phase slippage, a fact that is often overlooked. 
We also pointed out that the phase space volume is preserved under SBC when not considering 
radiation induced diffusion and damping. 

Following Ref. \cite{GuoXL}, which studied the horizontal closed orbit changes due to finite energy gains 
at rf cavities, we fully explored the problem by looking for a closed orbit in the 
4-dimension phase space, considering both finite energy gains at rf cavities and 
radiation energy losses in dipole magnets. We recovered the horizontal closed orbit change result in 
Ref. \cite{GuoXL} and found additional terms due to energy losses.

We carried out simulations with the accelerator modeling code AT~\cite{AccelTool} to 
verify the analysis. 
We found that the block diagonalization transformation matrix 
had high precision despite the approximations we made to get the analytic solution. 
The closed orbit results of AT also agreed well with the theory. 
%\appendix
%\section{}

% If you have acknowledgments, this puts in the proper section head.
\begin{acknowledgments}
The author thanks James Safranek, Andrei Terebilo and Jeff Corbett for the fruitful discussions 
with them which had significant influences over the development of this study. 
Discussions with Alex Chao helped the author on the history of the subject. 
This work was supported by Department of Energy
Contract No. DE-AC02-76SF00515.
\end{acknowledgments}

% Create the reference section using BibTeX:
\bibliography{xlcpaper}

\end{document}